\documentclass{elsart} 
\usepackage{amsmath}
\usepackage{graphicx}

\usepackage{supertab}

\newcommand{\bild}{8cm}
\begin{document}
\begin{frontmatter}

%\sloppy

%\thesaurus{00(0.0.0; 0.0.0; 0.0.0)} 

\title{Limits on the TeV flux of diffuse gamma rays as measured with the
HEGRA air shower array}

\collab{HEGRA Collaboration}

%\titlerunning{Upper Limit on the Diffuse Photon Background}
%\authorrunning{HEGRA Collaboration}

\author[heidelberg]{F.A. Aharonian},
\author[yerevan]{A.G.~Akhperjanian},
\author[madrid]{J.A.~Barrio},
\author[heidelberg]{K.~Bernl\"ohr},
\author[wuppertal]{H.~Bojahr},
\author[heidelberg]{O.~Bolz},
\author[kiel]{H.~B\"orst},
\author[madrid]{J.L.~Contreras},
\author[muenchen]{J.~Cortina},
\author[muenchen]{S.~Denninghoff},
\author[madrid]{V.~Fonseca},
\author[muenchen]{H.J.~Gebauer},
\author[madrid]{J.C.~Gonzalez},
\author[hamburg]{N.~G\"otting},
\author[hamburg]{G.~Heinzelmann},
\author[heidelberg]{G.~Hermann},
\author[heidelberg]{A.~Heusler},
\author[heidelberg]{W.~Hofmann},
\author[hamburg]{D.~Horns\thanksref{horns}},
\author[madrid]{A.~Ibarra},
\author[wuppertal]{C.~Iserlohe},
\author[heidelberg]{I.~Jung},
\author[heidelberg]{R.~Kankanyan},
\author[muenchen]{M.~Kestel},
\author[heidelberg]{J.~Kettler},
\author[heidelberg]{A.~Kohnle},
\author[heidelberg]{A.~Konopelko},
\author[muenchen]{H.~Kornmayer},
\author[muenchen]{D.~Kranich},
\author[heidelberg]{H.~Krawczynski},
\author[heidelberg]{H.~Lampeitl},
\author[madrid]{M.~Lopez},
\author[muenchen]{E.~Lorenz},
\author[madrid]{F.~Lucarelli},
\author[wuppertal]{N.~Magnussen},
\author[kiel]{O.~Mang},
\author[wuppertal]{H.~Meyer},
\author[muenchen]{R.~Mirzoyan},
\author[madrid]{A.~Moralejo},
\author[madrid]{E.~O\~na},
\author[madrid]{L.~Padilla},
\author[heidelberg]{M.~Panter},
\author[muenchen]{R.~Plaga},
\author[heidelberg]{A.~Plyasheshnikov\thanksref{altai}},
\author[hamburg]{J.~Prahl},
\author[heidelberg]{G.~P\"uhlhofer},
\author[kiel]{G.~Rauterberg},
\author[wuppertal]{W.~Rhode},
\author[hamburg]{A.~R\"ohring},
\author[heidelberg]{G.P.~Rowell},
\author[yerevan]{V.~Sahakian},
\author[kiel]{M.~Samorski},
\author[kiel]{M.~Schilling},
\author[wuppertal]{F.~Schr\"oder},
\author[kiel]{M.~Siems},
\author[kiel]{W.~Stamm},
\author[hamburg]{M.~Tluczykont},
\author[heidelberg]{H.J.~V\"olk},
\author[heidelberg]{C.~Wiedner} and
\author[muenchen]{W.~Wittek\thanksref{wow}}

\address[heidelberg]{Max-Planck-Institut f\"ur Kernphysik,
Postfach 103980, D-69029 Heidelberg, Germany}

\address[muenchen]{Max-Planck-Institut f\"ur Physik,
F\"ohringer Ring 6, D-80805 M\"unchen, Germany}

\address[madrid]{Universidad Complutense, Facultad de Ciencias
F\'{\i}sicas, Ciudad Universitaria, E-28040 Madrid, Spain} 

\address[hamburg]{Universit\"at Hamburg, II. Institut f\"ur
Experimentalphysik, Luruper Chaussee 149,
D-22761 Hamburg, Germany}

\address[kiel]{Universit\"at Kiel, Institut f\"ur Experimentelle und Angewandte Physik,
Leibnizstra{\ss}e 15-19, D-24118 Kiel, Germany}

\address[wuppertal]{Universit\"at Wuppertal, Fachbereich Physik,
Gau{\ss}str.20, D-42097 Wuppertal, Germany}

\address[yerevan]{Yerevan Physics Institute, Alikhanian Br. 2, 375036 Yerevan, 
Armenia}

\thanks[horns]{Now at  
Max-Planck-Institut f\"ur Kernphysik, Postfach 103980, D-69029 Heidelberg,
Germany}

\thanks[altai]{On leave from  
Altai State University, Dimitrov Street 66, 656099 Barnaul, Russia}

\thanks[wow]{Corresponding author : Wolfgang Wittek, 
e-mail address: wittek@mppmu.mpg.de}

%\mail{Wolfgang Wittek, \\Tel.: (Germany) +89 32354 291,\\

%\offprints{Wolfgang Wittek}

%\date{Received ; accepted }

%\maketitle

\newpage
\begin{abstract}
Using data from the HEGRA air shower array, taken in the period from April 
1998 to March 2000, upper limits on the ratio $I_\gamma /I_{CR}$
of the diffuse photon flux $I_\gamma$
to the hadronic cosmic ray flux $I_{CR}$ are determined
for the energy region 20 TeV to 100 TeV. The analysis uses a
gamma-hadron discrimination which is based on differences in the
development of photon- and hadron-induced air showers after the shower maximum.
A method which is sensitive only
to the non-isotropic component of the diffuse
photon flux yields an upper limit of 
$I_\gamma /I_{CR}\;(\rm{at}\;54\;\rm{TeV})\;<\;2.0\times10^{-3}$ (at the 90\% 
confidence level) for a sky region near
the inner galaxy ($20^\circ <$ galactic longitude $<\;60^\circ$ and
$|$galactic latitude$|\;<\;5^\circ$). A method which is sensitive to both the
isotropic and the non-isotropic component yields global upper limits of
$I_\gamma /I_{CR}\;(\rm{at}\;31\;\rm{TeV})\;<\;1.2\times10^{-2}$ and 
$I_\gamma /I_{CR}\;(\rm{at}\;53\;\rm{TeV})\;<\;1.4\times10^{-2}$ (at the 
90\% confidence level). 

%\keyword{Cosmic rays, Diffuse gamma-ray Galactic emission; 
%Gamma rays: observations}

%\noindent{\it Keywords}: Cosmic rays, Diffuse gamma-ray Galactic emission; 
%Gamma rays: observations

\begin{keyword}
Cosmic rays, Diffuse gamma-ray Galactic emission; Gamma rays: observations.
\end{keyword}

\end{abstract}
\end{frontmatter}

%\twocolumn

\section{Introduction}
The flux of the diffuse photon background radiation, 
its spectrum and its evolution
with time are very interesting topics of astrophysics. In nearly all 
energy regions both an extragalactic and a galactic component of the diffuse
radiation can be identified. The former component, which is essentially
isotropic, gives information about processes and developments at large
distances from our galaxy, ranging partly back to the early universe. Details
about our own galaxy like the interstellar matter, radiation and magnetic
fields as well as about
the origin and the propagation of galactic cosmic rays can be
deduced from investigations of the galactic diffuse emission.

Direct measurements of the diffuse emission in the 30 MeV to 50 GeV energy 
range are provided by the EGRET experiment \cite{hunter,sreekumar}.
At high galactic latitudes an extragalactic component has been observed,
a significant fraction of which is attributed to the direct emission from AGNs
\cite{chiang}. The main part of the diffuse emission is concentrated at
the galactic disc and the inner galaxy, which therefore points to a galactic 
origin. The EGRET data below 1 GeV 
can be well explained by interactions of cosmic-ray electrons and
protons with the interstellar radiation field and
with the interstellar matter \cite{bertsch}. The
dominant interaction processes are $\pi^0$ production by nucleon-nucleon
interactions, inverse Compton (IC) scattering with low-energy photons and
high-energy electron bremsstrahlung. Above 1 GeV the observed photon flux
exceeds the model predictions \cite{hunter}. A possible explanation for this
discrepancy is that the average interstellar proton spectrum is harder than
the spectrum observed in the local neighbourhood, leading to a harder
$\pi^0$ and thus also to a harder diffuse gamma-ray spectrum 
\cite{mori,gralewicz,moskalenko98,aharonian00}. The discrepancy may also
be due to an underestimation of the
inverse-Compton contribution, which starts to become relevant above 1 GeV 
\cite{porter,pohl,porter1,strong,dar,strong1,moskalenko1}. 
The favoured hypothesis for an enhanced IC
contribution is an average electron spectrum in the galaxy which is
substantially harder than that measured locally
\cite{pohl,strong1,moskalenko1}. An alternative or additional
contribution could be gamma-radiation from supernova remnants and
pulsars \cite{berezhko}. 

It is interesting to note that in models with increased inverse-Compton
contribution \cite{dar,strong1} not only the excess of photons 
above 1 GeV can be explained but also the longitude and
latitude profiles up to the galactic poles can be reproduced. This could
be relevant for the experimental determinations of the isotropic
extragalactic emission.

In these models the IC contribution to the diffuse photon background
dominates above 1 GeV. Depending on the cutoff energy for the electron
injection spectrum a maximum of the differential photon flux 
(multiplied by $E^3$) is predicted between 30 TeV and 200 TeV \cite{porter1}. 
For the inner galaxy the flux at the maximum
corresponds to a fraction of $\sim 5\times 10^{-4}$ of the cosmic ray flux
(see Fig. \ref{fig15}).
The IC contribution is rather broad in galactic latitude and extends up
to the galactic poles, and even there it is comparable to or greater than
the expected isotropic extragalactic contribution \cite{dar,strong1}.

The broad distribution in galactic latitude is due to the broad distribution
of the interstellar radiation field, which is the target field for the 
IC process.
In contrast, a model with enhanced $\pi^0$ production would predict a
gamma radiation which is more confined to the galactic disc, because of the
narrow distribution of the interstellar matter, the target for
the $\pi^0$ production process. This difference may allow one to 
distinguish between the two hypotheses.

Contributions to the extragalactic diffuse emission below 100 TeV are
expected from the cascading
of ultra-high-energy photons and electrons on the Cosmic Microwave Background
(CMB), creating
eventually a pile-up of gamma-rays in the 10 - 100 TeV region
\cite{hill,berezinsky}.
The ultra-high-energy photons and electrons may be decay 
products of supermassive particles \cite{sigl}, themselves
radiated during collapse or annihilation of topological defects
\cite{aharonian2}. Alternatively they may result from
the interaction of ultra-high-energy hadronic cosmic rays with the CMB 
\cite{halzen}. The expected level of gamma flux
contributed by the cascading process  is estimated to be 10$^{-5}$ of the
cosmic ray flux \cite{halzen,protheroe,sigl1}.

In the 1 TeV to 1 PeV energy range there are two experiments which claim
the observation of a diffuse photon signal:
\cite{nikolsky} and \cite{suga}. The result reported by \cite{nikolsky} is
only a 2.8 sigma effect and may therefore well be considered as an upper
limit. Moreover, there
may be an additional uncertainty of the result due to uncertainties on the
muon content of photon showers \cite{drees}. The gamma-ray excess reported
by \cite{suga} is a 3.8 sigma effect, which may be correlated with the Loop 1
region. In a later publication by the same collaboration \cite{kakimoto}
only an upper limit of the diffuse gamma-ray flux is given.
All other experiments in the 1 TeV to 1 PeV energy region
give upper limits on the diffuse photon flux
(see Tables \ref{table2} and \ref{table5} and Figs. \ref{fig15} and 
\ref{fig16}).

In the present analysis \cite{denninghoff}
an independent new determination of an upper limit on the diffuse
photon background is performed in the 20 TeV - 100 TeV region, where only
two other experiments have reported results \cite{he,sasano}, in
addition to previous measurements with the HEGRA array \cite{karle1,schmele}.

The energy region below and around 100 TeV is of particular interest because 
on the one hand the extragalactic component of the diffuse photon background
is expected to contribute predominantly in this energy region. On the other 
hand, according to some models \cite{porter1}, the ratio $I_\gamma / I_{CR}$
of the differential diffuse photon flux $I_\gamma$
of galactic origin to the cosmic ray flux $I_{CR}$
has a maximum in the 100 TeV region (see Figs. \ref{fig15} and 
\ref{fig16}).

The paper is structured as follows: Section 2 describes the experimental
details, Section 3 the Monte Carlo simulation. In Section 4 the
definition of the data sample to be used in the analysis is given. The
measurement of the upper limit on the diffuse photon flux is presented in
Section 5 and the results are discussed in Section 6.
 
\section{Experimental Details}
\subsection{The Detector}
The data were taken with the HEGRA air shower array located at 2200 m
a.s.l. on the Canary island La Palma. Since April 1998 the array consisted
of 97 wide-angle Cherenkov detectors (AIROBICC) and of 182 scintillation
detectors. A fire in October 1997 had destroyed 39 AIROBICC detectors, all 
of which were rebuilt, and 65 scintillation detectors, of which only 4 were 
rebuilt. The two types of detectors measure the Cherenkov radiation and
the secondary particles (mainly photons, converted in a 5 mm thick lead 
plate above the $4\;{\rm cm}$ thick plastic scintillator sheet, and electrons) 
respectively, which are produced in air showers induced by primary cosmic 
rays (including gammas) in the atmosphere.
The AIROBICC detectors were distributed over an area of
$200\;\mathrm{m}\times 200\;$m. This area contained completely the 
$150\;\mathrm{m}\times 200\;$m area covered by the scintillator array. The
detectors are described in detail in \cite{merck} \mbox{and \cite{karle}.} 

The experimental trigger required a time coincidence within 150 ns of $\geq$
6 Cherenkov detectors (AIROBICC trigger) or of $\geq$ 14 scintillation
detectors (scintillator trigger). The trigger rate was 28 Hz on the
average. 

The data used in this analysis were taken in the period from April 1998 to
March 2000. Only data with an AIROBICC trigger have been considered.
The total effective observation time amounts to $\sim 1464$ hours.

\subsection{Detector Calibration and Shower Reconstruction}
The detectors register the arrival time ({\it time} data) and the 
light flux ({\it amplitude} data) hitting the photocathode of the 
photomultipliers.
For each detector the amplitude is given in terms of the number of ADC 
channels. In the case of the scintillator array the relative calibration 
of amplitudes
between the individual detectors is done on the basis of the so-called
'1-MIP peak'. The 1-MIP peak is the position of the maximum in the ADC
spectrum corresponding to the detector response for 1 minimum ionizing
particle \cite{moralejo}. For the Cherenkov array
a relative calibration between the individual detectors is achieved by
normalizing the high-amplitude tails of the individual ADC spectra to each 
other \cite{moralejo}. The absolute calibrations of the amplitudes are 
described in Section 3.2.

By the absolute calibrations the amplitudes are transformed into a density
of Cherenkov photons impinging on the surface of the
Cherenkov detectors or a density of secondary electrons and photons
impinging on the surface of the scintillation detectors
respectively.

For each shower the shower direction, the position of the shower core and 
the Cherenkov light density $\rho(r)$ as a function of the distance $r$ 
from the shower core were reconstructed by a simultaneous fit 
$(global\;AIROBICC\;fit)$ to the $time$ and $amplitude$ data from the 
Cherenkov detectors. From 
$\rho(r)$ the Cherenkov light density $L_{90}$ at $r\;=\;90$ m and the
light radius $R_L\;=\;-1\;/\;({\rm d}ln\rho(r)/{\rm d}r)$ in the region 
50 m $<\;r\;<$ 120 m were derived.
The total number $N_s$ of photons and electrons at the 
detector level was determined by a maximum likelihood fit to the particle
densities from the scintillation detectors.

$L_{90}$ is known to be a good measure of the shower energy.
In the following $L_{90}$ and $R_L$ are assumed to be given in units of
\mbox{[no.of photons / $\rm{m}^2$]} and [m] respectively.

The quality of the various fits to the $time$ and $amplitude$ data, 
expressed in terms of 
$\chi^{2}-$probabilities and in terms of errors of fitted parameters, 
is used below 
in the definition of the selection criteria for the showers.  

\section{Monte Carlo Simulation and Data-Monte Carlo Comparison}
\subsection{Monte Carlo Simulation}
An essential part of the present analysis is the Monte Carlo (MC) 
simulation of the experiment. The simulation comprises :
\begin{itemize}
\item[--]The simulation of the physical processes in the shower
development, including the production of Cherenkov photons. The CORSIKA
program, version
4.068, was used for this step \cite{capdevielle}. 
\item[--]The simulation of the detector. A comprehensive discription is
given in \cite{martinez}. 
\item[--]The simulation of the chemical composition and of the energy
spectrum of the charged cosmic rays. The simulation was based on the
compilation \cite{wiebel} (see Table \ref{table1}). 
\end{itemize}

MC data were generated for primary photons, protons, He-, O- and Fe nuclei,
for the zenith angles $\Theta$ = 10$^\circ$, 20$^\circ$ and 30$^\circ$ and
for the energy region 5~TeV $<E<1000$~TeV. For each particle type and each
zenith angle approximately the same number of showers was generated, giving
a total of 133 000 showers. By assigning to each of these showers 20
different core positions, chosen randomly in a $300\;\mathrm{m}\times
300\;$m area centered at the actual detector array, the statistics of 
quasi-independent showers was increased by a factor of 20.

The generated showers were given appropriate weights to simulate the
expected distributions in E, $\Theta$ and the cosmic ray species.
For the comparisons with the experimental data an admixture of
photons with $\phi_0 = 0.0002582\; (\rm m^{2}\cdot s\cdot
sr\cdot \rm TeV)^{-1}$ has been assumed in the MC sample
(see Table \ref{table1}), corresponding to 1/1000 of the hadronic cosmic
ray flux at 1 TeV.

In the following 'MC events' always means 'weighted MC events'.

\subsection{Absolute Calibration of $N_s$ and $L_{90}$}
The absolute calibration of $N_s$ and $L_{90}$  
is done separately for each subrun (corresponding to 30 minutes
of observation time).
The calibration factor for $N_s$ is determined such that
\mbox{$\langle\log_{10}(N_s)\rangle$} in the experimental data agrees with
\mbox{$\langle\log_{10}(N_s)\rangle$} in the MC data. This is done using 
samples defined 
by the standard selection criteria (see Section 4).

The absolute $N_s$ calibration defines - through the MC simulation - the
absolute energy scale in the experimental data. By comparing with other
methods of calibrating $N_s$, for example using the 1-MIP peak not only for
the relative but also for the absolute calibration of $N_s$ \cite{moralejo}, 
the error in the absolute energy scale is
estimated to be in the order 20\%.
 
After calibration, the $\log_{10}(N_s)$ distributions of the experimental
data and of the MC data agree fairly well (see Fig. \ref{fig1}). There is also
reasonable agreement  in the total number of showers: for a small subset of
runs the total number of showers in the experimental data was compared with
the corresponding MC number expected on the basis of the measured on-time
and the all-particle cosmic ray flux from \cite{wiebel}. There is
agreement within 7\%. This can be regarded as a good consistency between
experimental data and MC data, given the fact that already a miscalibration
of $N_s$ by 10\% would yield a discrepancy in the number of reconstructed
showers of ($1.10^{1.7} - 1.0$) = 17.6\%.

The absolute calibration of $L_{90}$ is done using the calibrated $N_s$
values, namely by determining a calibration factor for $L_{90}$ such that
$\langle\log_{10}(N_s/L_{90})\rangle$ in the experimental data coincides with
$\langle\log_{10}(N_s/L_{90})\rangle$ in the MC data. The calibration factor 
is determined
separately for each of five bins in the variable $(1/R_L)$.

\subsection{Adjustment of Errors in the Monte Carlo Simulation}
A comparison of the fluctuations of arrival times and amplitudes between
experimental data and MC data gave good agreement for the data from the
scintillation detectors. For the Cherenkov detectors the fluctuations were
found to be significantly lower in the MC data than in the experimental
data. The difference is explained on the one hand by an underestimation of 
the average light of the night sky background in the MC simulation. 
On the other hand, the difference may be due to subtle effects in the
conversion chain from the photons to the final electronic signal, such as
variations in the Winston cone reflectivity, non-uniformity in the
quantum efficiency and the photoelectron collection efficiency of the
photomultiplyers, etc. . Instead of repeating the MC simulation with a more 
realistic night sky background and with a refined simulation of the
detector non-uniformities, the
fluctuations of arrival times and amplitudes from the Cherenkov detectors
were randomly increased in the MC-generated data to agree with those
in the experimental data. On the average the RMS of the Cherenkov photon 
densities and that of the Cherenkov photon arrival times was increased by
a factor of 1.8.

\subsection{Data-MC Comparison}
\begin{figure}[p]
\includegraphics[clip,width=\bild]{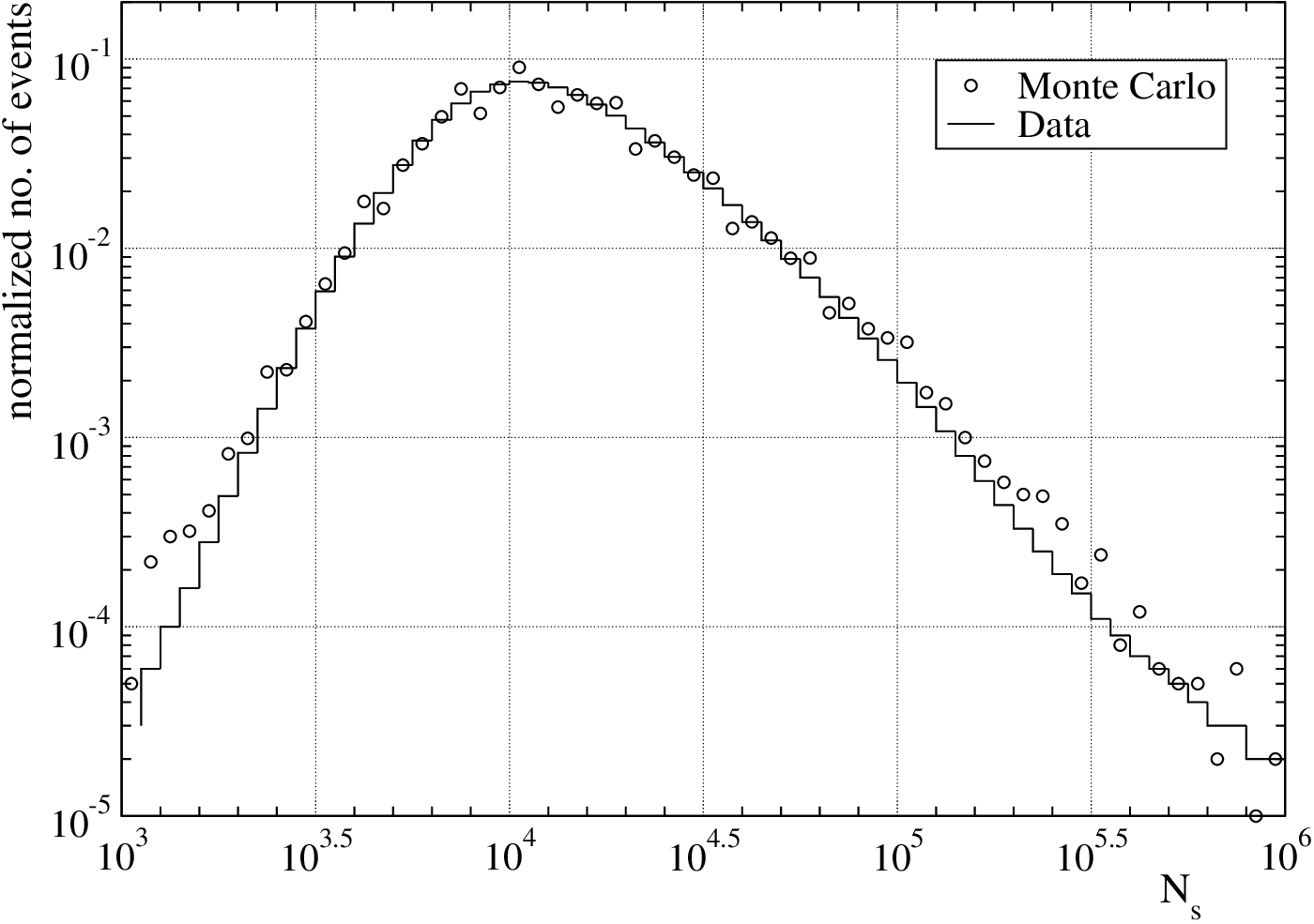}
\caption{Distribution of $\log_{10}(N_s)$ for the experimental data 
(histogram) and for the MC data (open circles).}
\label{fig1}
\end{figure}

\begin{figure}[p]
\includegraphics[clip,width=\bild]{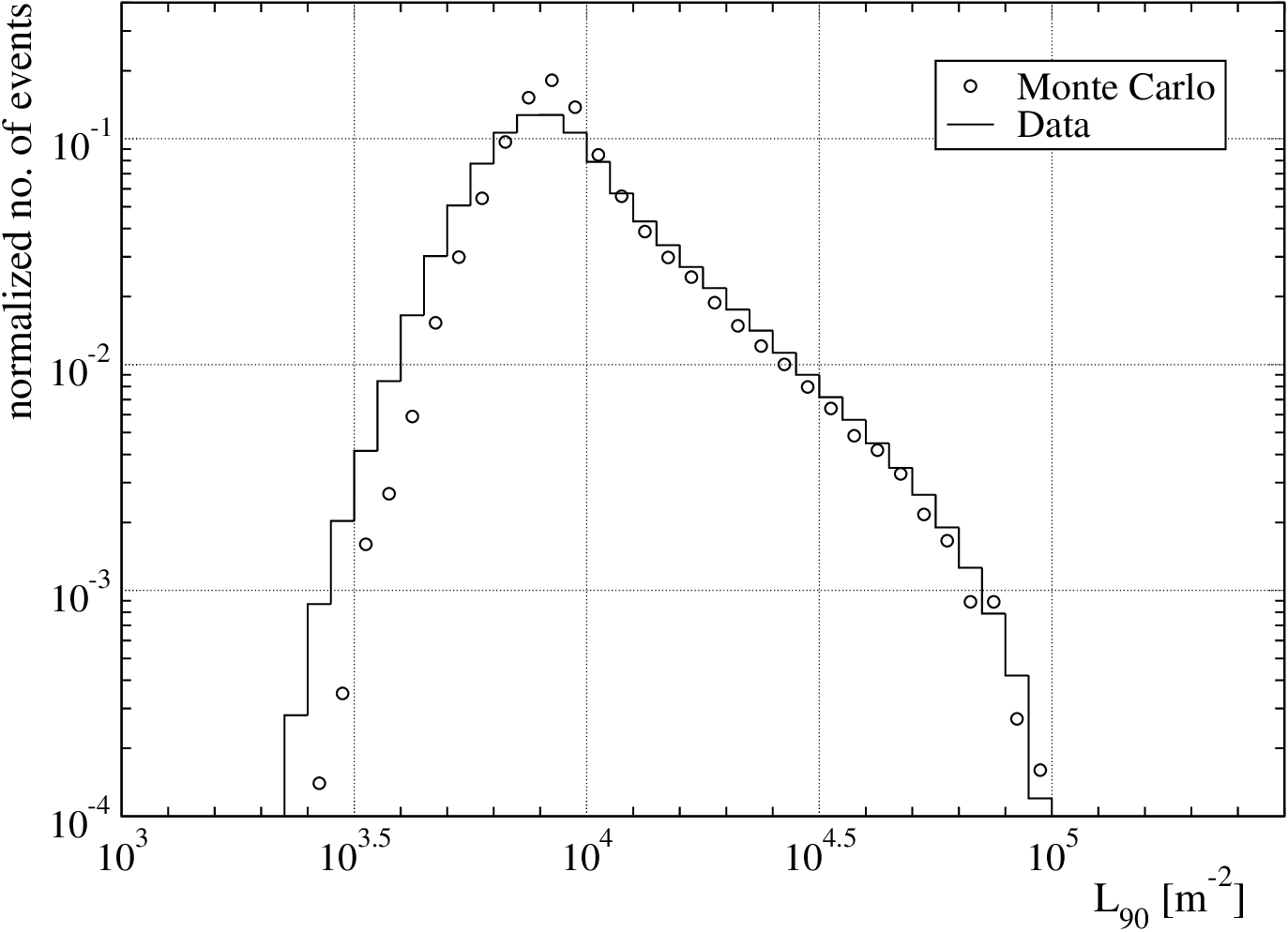}  
\caption{Distribution of $\log_{10}(L_{90})$ for the experimental data 
(histogram) and for the MC data (open circles).}
\label{fig2}
\end{figure}

\begin{figure}[p]
\includegraphics[clip,width=\bild]{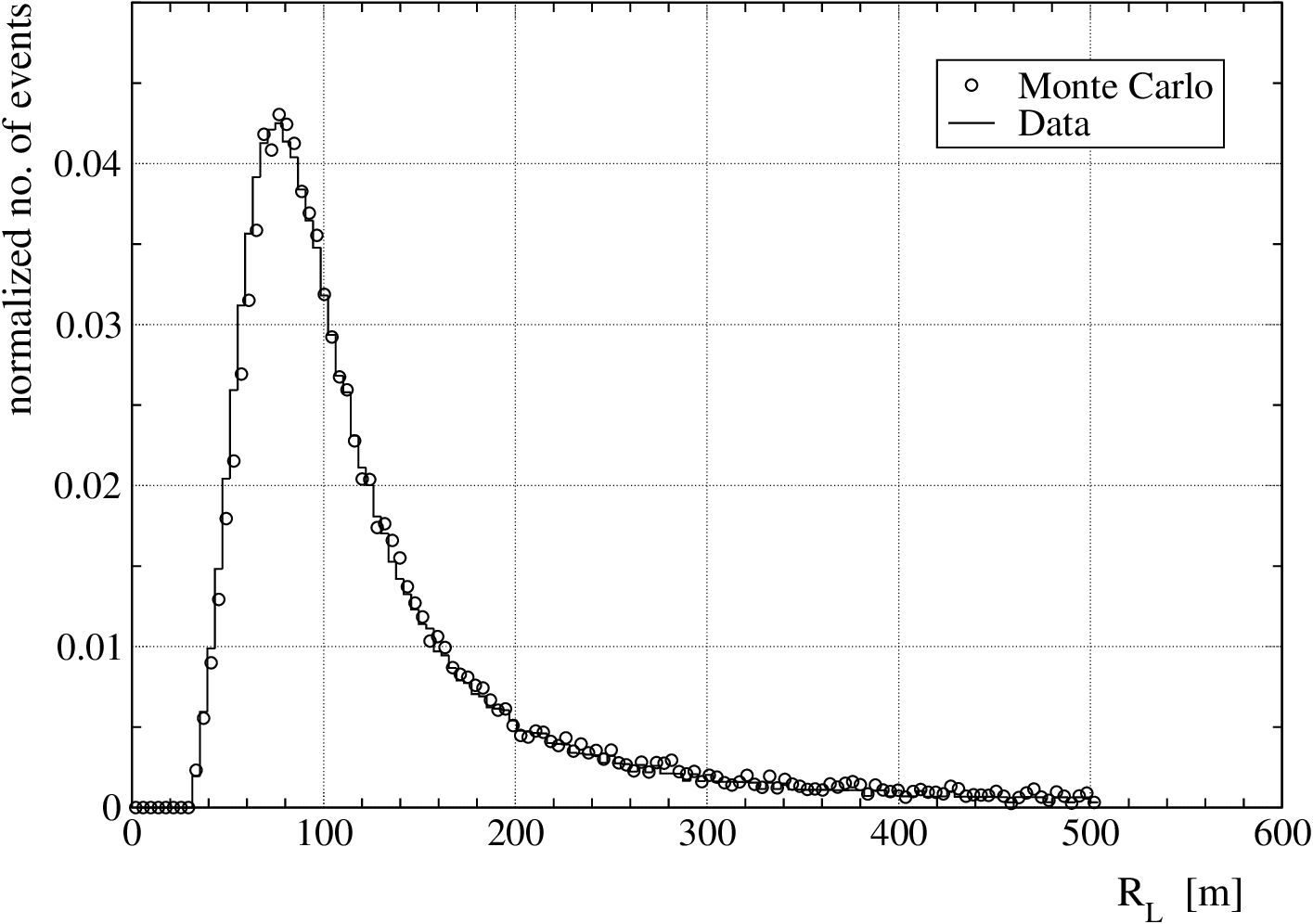} 
\caption{Distribution of $R_L$ for the experimental data (histogram)
and for the MC data (open circles).}
\label{fig3}
\end{figure}

Some comparison of experimental data and MC data are shown in
Figs. \ref{fig1} to \ref{fig8}. The data samples are defined by the standard
selection criteria listed in Section 4. Because these selection criteria
are tighter than the trigger condition no simulation of the trigger was
necessary. In each of the Figs. \ref{fig1} to \ref{fig3}
the two distributions were normalized to the same total number of events.
For $\log_{10}(N_s)$, $R_L$ and also for $\log_{10}(N_s/L_{90})$ versus
$(-1/R_L)$ the agreement between data and MC is quite good. 
A large discrepancy is found for
$\log_{10}(L_{90})$ at $\log_{10}(L_{90}) < 4.0$. This may be due to the time
dependence of the light of the night sky in the experimental data which
was not taken into account in the MC simulation.  

\subsection{Comparison of photon-induced to hadron-induced showers
 in the MC data}
An experimental upper limit of the flux of diffuse photons will obviously
be the tighter the better photons can be discriminated from hadrons. It is
therefore worthwhile searching for measurable quantities which allow a
photon-hadron discrimination at least to a certain degree.
In the present analysis this is done exclusively with MC data: As no prominent 
photon signal  has been found so far
with HEGRA-array data there is no experimental data sample, sufficiently
enriched with photons, that could be used for studying the photon-hadron
separation.

As shown in \cite{arqueros} useful variables in this 
context are $1/R_L$ and the ratio \\
$\log_{10}(N_s/L_{90})$: because
hadron-induced showers in general develop more slowly after the shower
maximum than photon-induced showers, the ratio $\log_{10}(N_s/L_{90})$ at
fixed position of the shower maximum (estimated by $1/R_L$ \cite{patterson})  
is expected 
to be on the average larger for hadron than for photon showers. This
behaviour is illustrated in Fig. \ref{fig9}. The photon-hadron difference is
quite independent of $\ln(L_{90})$ and $\Theta$ (not shown). 
In Fig. \ref{fig8} the
experimental data are compared with the MC-(hadron + photon) sample. The
agreement for the average $\log_{10}(N_s/L_{90})$ at fixed $(-1/R_L)$ is not
surprising because it is the result of the absolute calibrations, 
described in Section 3.2.

MC studies have shown that the shape parameters of the lateral particle
distribution, or alternatively the age parameter in the NKG function,
as determined in the fit to the $amplitude$ data from the scintillation
detectors,
have a potential for discriminating between photons and hadrons which may
be superior to that of the variable $1/R_L$. An approach in this direction
was followed in \cite{prahl}. In the present analysis, however, this
potential could not be exploited due to the unsatisfactory results of the 
comparison between experimental data and MC data : no good agreement was
found in the distribution of the age parameter, the distribution being wider 
and shifted towards lower values in the MC data.

\section{Data Sample}
The aim of the selection criteria 
is to select well reconstructed showers, to concentrate on
energies close to the threshold of the HEGRA-array and to retain only data
taken under good observation conditions. This is achieved by the set of 
requirements (called `standard selections' in the following):

\begin{itemize}
\item[--]$\chi^{2}$-probability of the global AIROBICC fit :
$P(\chi^2)\;>\;1\%$

\item[--]zenith angle $\Theta < 35^\circ$

\item[--]$-0.03\; \rm{m}^{-1} < (-1/R_L) < 0.01\; \rm{m}^{-1}$

\item[--]fitted errors of the zenith angle $\Theta$ and of the 
azimuthal angle $\phi$ : \\ 
$\Delta\Theta < 0.4^\circ$,       $\sin\Theta\cdot\Delta\phi < 0.4^\circ$

\item[--]fitted error on $\rm x$ and $\rm y$ position of the shower core :\\
$\Delta \rm x_{core} < 15\; m,\;\Delta \rm y_{core} < 15\; m$

\item[--]fitted error on $\ln(L_{90})$ : $\Delta\ln(L_{90})\;<\;0.3$ 

\item[--]$\chi^{2}$-probability of the maximum likelihood fit to the lateral
particle distribution $(\rho_s(r))$ from the scintillation detectors : 
$P(\chi^2)\;>\;5\%$

\item[--]fitted error on the total number of particles at the detector  
level :\\
$\Delta \log_{10}(N_s) < 0.4$  

\item[--]rejection of very-high-energy showers: \\   
$\ln(L_{90}) < 11.0 + 0.45 \dfrac{\cos\Theta - \cos(30^\circ)} 
                                {\cos(10^\circ) - \cos(30^\circ)}$ 

This cut eliminates photon showers above $\sim$~320 TeV, proton showers
above $\sim$~640 TeV and Fe showers above $\sim$~1000 TeV. 
The cut has to be applied in order to ensure consistent conditions for
experimental data and MC data, which were generated only up to energies of
1000 TeV.
\end{itemize}

Out of the 163 million triggered showers 19.3 million pass the standard 
selections. The number of triggered showers (100\%) is reduced to 
56\%, 36\% and 12\% by requiring one after another a successful global 
AIROBICC fit, a successful fit to the lateral particle distribution 
$\rho_s(r)$ and well reconstructed showers. The strong reduction is explained
by the large experimental errors in showers which are close to the
energy threshold of the trigger.
 
\begin{figure}[p]
\includegraphics[clip,width=\bild]{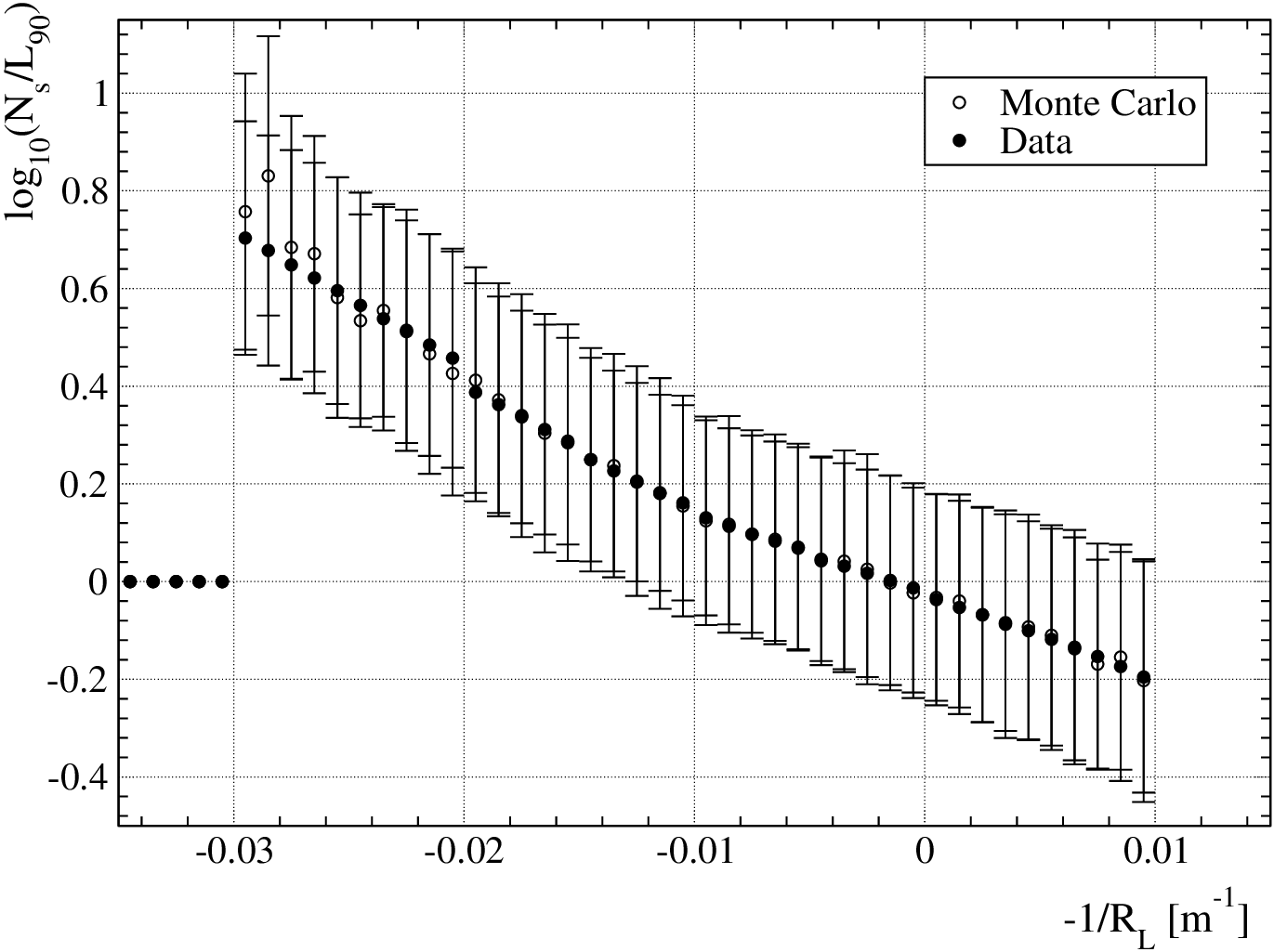} 
\caption{Average value of $\log_{10}(N_s / L_{90})$ as a function of $(-1/R_L)$
for the experimental data (full circles) and for the MC data (open
circles). The error bars represent the RMS of the $\log_{10}(N_s / L_{90})$
values in each bin of $(-1/R_L)$.}
\label{fig8}
\end{figure}

The main characteristics of the data sample defined by the standard
selections are (see also Figs. \ref{fig1} to \ref{fig8}):

\begin{itemize}
\item[--]average fitted errors  :\\
\begin{tabular}{ll}
$\langle\Delta \rm x_{core}\rangle\approx\langle\Delta \rm y_{core}\rangle = 4.1\; m$\;\;\;\;\;  &
$\langle\Delta\Theta\rangle\approx \langle\sin\Theta\cdot\Delta\phi\rangle = 0.12^\circ$  \\
$\langle\Delta \ln(L_{90})\rangle$ = 0.12  &
$\langle\Delta \log_{10}(N_s)\rangle$ = 0.12  \\
$\langle\Delta (1/R_L)\rangle = 0.003$ \rm m$^{-1}$ &
\end{tabular}

\item[--]average zenith angle $\langle\Theta\rangle = 20^\circ$

\item[--]average $\ln(L_{90})\;=\;9.16\;\;\;\;\; 
(L_{90}$ in units of no.of photons / \rm m$^{2}$)

\item[--]average $\log_{10}(N_s)$ = 4.12  \\
\end{itemize}

Applying the standard selections to the MC data one finds :

\begin{itemize}
\item[--] the energy region for reconstructed photon showers  
 (10\% and 90\% quantiles) is $20 \rm\; TeV < E_\gamma < 
100 \rm\; TeV$,  with an average energy of 53 TeV  \\

\item[--]the energy region for reconstructed hadron showers  
 (10\% and 90\% quantiles) is $40 \rm\; TeV < E_h < 210 \rm\; TeV$,  with an average energy of 110 TeV  \\

\item[--]the ratio of the number of reconstructed photon showers to the 
number of
reconstructed hadron showers $N_\gamma / N_h$ is 1/346. Note that the
$\gamma / h$ flux ratio assumed in the MC simulation was 1/1000 at 1 TeV
(see Table \ref{table1}). This apparent suppression (by a factor of $\sim 3$)
of hadron showers is due to the fact that with respect to $L_{90}$ photon 
showers of a given energy resemble hadron showers of about twice the energy, 
where the integrated flux is reduced by a factor of $2^{1.7} \approx 3$.

\end{itemize}

The effective collection area as a function of the energy is shown in Fig.
\ref{fig0} for photon, proton and Fe showers. Defining the threshold energy
as the energy where the effective collection area reaches 50 $\%$ of its
maximum value one obtains for photons a threshold energy of 35 TeV, if one
averages over all zenith angles, and 22, 28 and 45 TeV for the zenith
angles 10$^\circ$, 20$^\circ$ and 30$^\circ$ respectively. 

As can be seen from Figs. \ref{fig1} to \ref{fig8} the average values of
the different quantities are similar for experimental data and for MC data.
For MC events it has been verified that the average fitted error of
$\Theta$ is close to the RMS of 
($\Theta_{fitted} -\Theta_{generated}$). Similar statements hold for 
other quantities. The value of the average fitted error
$\langle\Delta\Theta\rangle\approx0.12^\circ$ therefore implies
that the angular resolution for
the data sample defined by the standard selections is around $0.12^\circ$,
not including possible systematic errors.

In the subsequent analysis the standard selections have always been applied.

\section{Determination of Upper Limits on the Flux of Diffuse Photons}
\begin{figure}[p]
\includegraphics[clip,width=\bild]{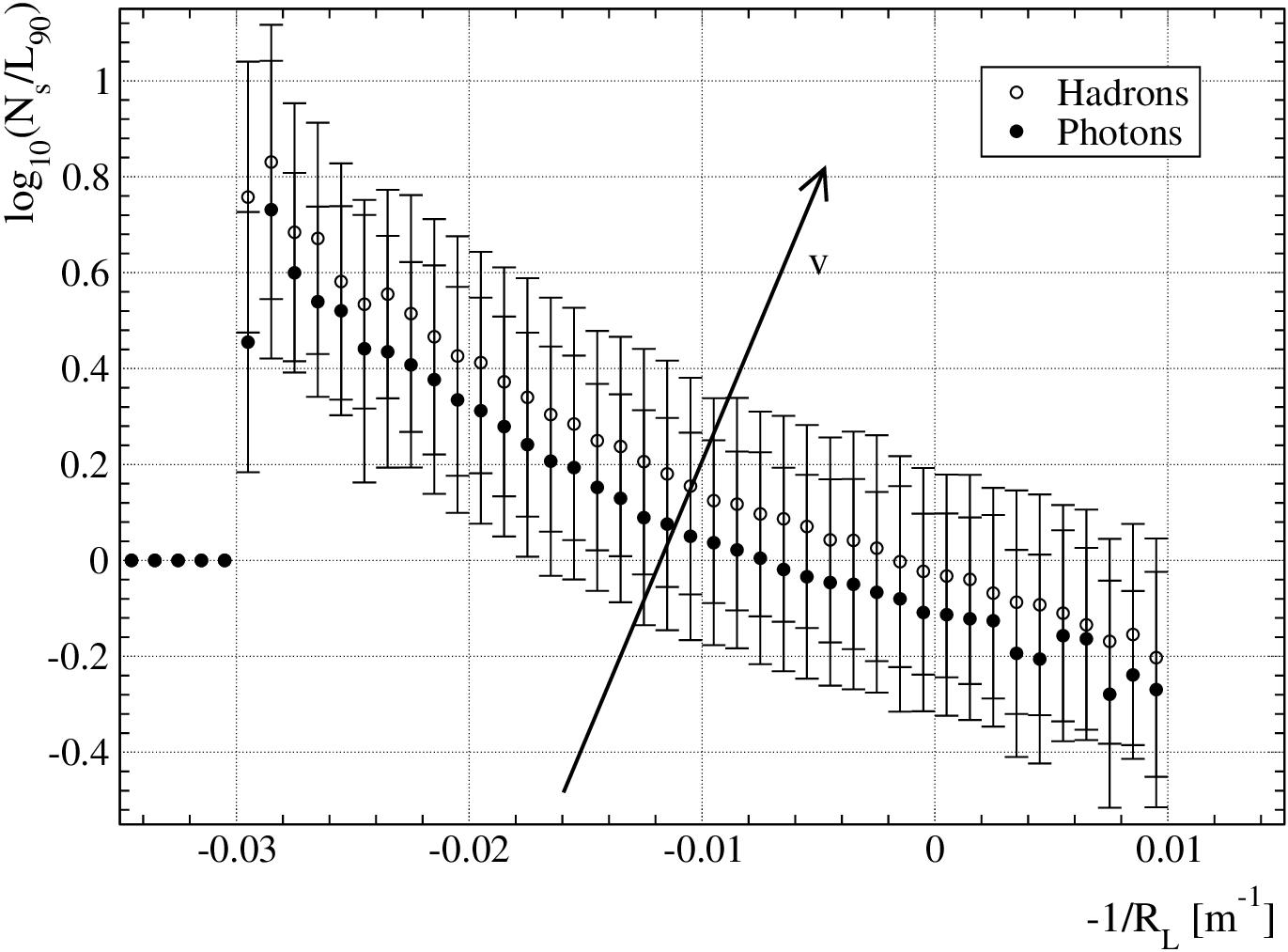} 
\caption{Average value of $\log_{10}(N_s / L_{90})$ as a function of $(-1/R_L)$
for photon-induced (full circles) and hadron-induced (open circles)
showers. The error bars represent the RMS of the $\log_{10}(N_s / L_{90})$ 
values in each bin of $(-1/R_L)$. The arrow indicates the axis of the
variable $v$ defined in the text.}
\label{fig9}
\end{figure}

\begin{figure}[p]
\includegraphics[clip,width=\bild]{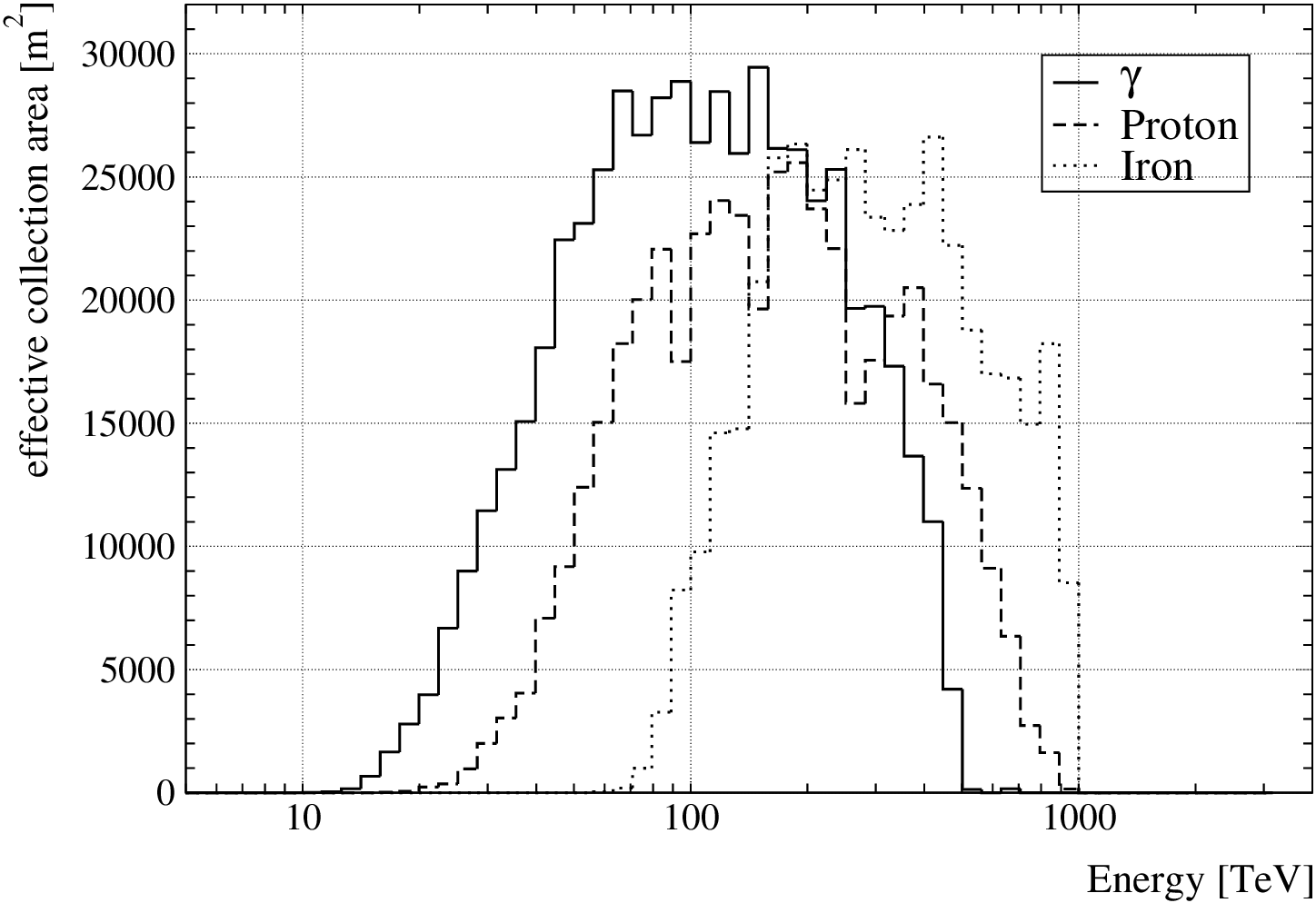}
\caption{Effective collection area as a function of the energy for 
MC showers, after applying the standard selections and averaging over all
zenith angles, for photon (solid histogram), proton (dashed histogram) and
Fe showers (dotted histogram).}
\label{fig0}
\end{figure}

As shown in Section 3.5 some difference between photon and hadron 
showers is seen in the plot of $\log_{10}(N_s/L_{90})$ versus $(-1/R_L)$
(Fig. \ref{fig9}). Under all projections of the 2-dimensional distribution
$\log_{10}(N_s/L_{90})$ versus $(-1/R_L)$ onto some axis, the one onto the
$v-$axis, indicated by the arrow in Fig. \ref{fig9}, can be expected to yield
a relative small overlap between the distributions for photons and hadrons.
The distribution in the variable $v$, defined as 
\vspace*{-3.0ex}
\begin{eqnarray}
v &= &\log_{10}(N_s/L_{90}[{\rm m}^{-2}]) + 23.43 \cdot 
(0.005 - 1/R_L[{\rm m}])\;\;\;\;, 
\end{eqnarray}
will therefore be sensitive to the relative contribution of photons and
hadrons.

In the following the distributions of the variable $v$ will be exploited for
determining upper limits of the diffuse photon background. Two approaches 
will be followed

\begin{itemize}
\item[-] Determination of a global upper limit by comparing the experimental
distribution of $v$ with the MC expectations for
primary photons and hadrons (Sect. 5.1). \\

\item[-] Determination of an upper limit for a specific region in the galaxy
by comparing the experimental distributions of $v$ in this region with that
of another region in the galaxy (Sect. 5.2).
\end{itemize}

\subsection{Global upper limit of $I_\gamma / I_{CR}$}
The distributions in $v$ are denoted as 

\begin{tabular} {llll}
    $q_{data}$ &= &$dN_{data} / dv$ 
               &for reconstructed showers of the experimental data  
               \\

    $q_\gamma$ &= &$dN_{MC}^\gamma / dv$ 
               &for reconstructed MC-photon showers  \\

    $q_h$      &= &$dN_{MC}^h / dv$ 
               &for reconstructed MC-hadron showers \\
    \end{tabular}

Both $q_h$ and $q_\gamma$ are normalized such that they correspond to the
absolute distributions of reconstructed MC showers for
an effective on-time of 1 s, assuming for photons and 
hadrons the same differential flux at 1 TeV of 
0.2582 particles / ($\rm m^{2}\cdot s\cdot sr \cdot TeV$), which is the
measured differential flux of hadronic cosmic rays. 

With fixed spectral indices of the photon and hadron fluxes 
(Table \ref{table1}) the contribution from photons to the cosmic ray flux 
may be specified by the ratio $r_{0} = I_\gamma^0 / I_{CR}^0$ of the
differential photon flux and the differential hadron flux at some arbitrary
energy $E_0$. $E_0$ is chosen as 1 TeV.

The ratio $r_0$ is then determined by fitting
$q_{data}$ by a superposition of $q_\gamma$ and $q_h$ :
\vspace*{-3.0ex}
%\begin{tabular}{lll}
% $q_{data}$ &= &$T_{on}[s] \cdot a\cdot (q_h + r_0\cdot q_\gamma)$
%\end{tabular}
\begin{eqnarray}
 q_{data} &= &T_{on}[{\rm s}] \cdot a\cdot (q_h + r_0\cdot q_\gamma)
 \label{eq4}
\end{eqnarray}
$T_{on}$ is the total on-time of the experimental data, and $a$ and
$r_0$ are free parameters to be determined in the fit. 
 
Using the spectral indices adopted in the MC simulation (see 
Table \ref{table1}) $r_0$ can be converted into a ratio $r_1$ at any other
energy $E_1$ (see below). If one wants to determine in the fit $r_1$
instead of $r_0$, the term $r_0\cdot q_{\gamma}$ in eq. (\ref{eq4}) has to
be replaced by $r_1\cdot q_{\gamma,1}$ with $q_{\gamma,1}=
q_{\gamma}\cdot r_0/r_1$. The term $q_h$ would remain unchanged, 
because $q_h$ has been normalized to the measured differential flux
of hadronic cosmic rays already. The choice of the reference energy $E_0$ is 
therefore completely arbitrary.

In case of a perfect MC simulation and an accurate measurement of the
on-time $T_{on}$ the normalization factor $a$ should come out to be 1.
Leaving it free in the fit makes the result for $r_{0}$ independent of
small inconsistencies in normalization between experimental data and MC
data and compensates for small deficiencies of the MC simulation.

\begin{figure}[p]
\includegraphics[clip,width=\bild]{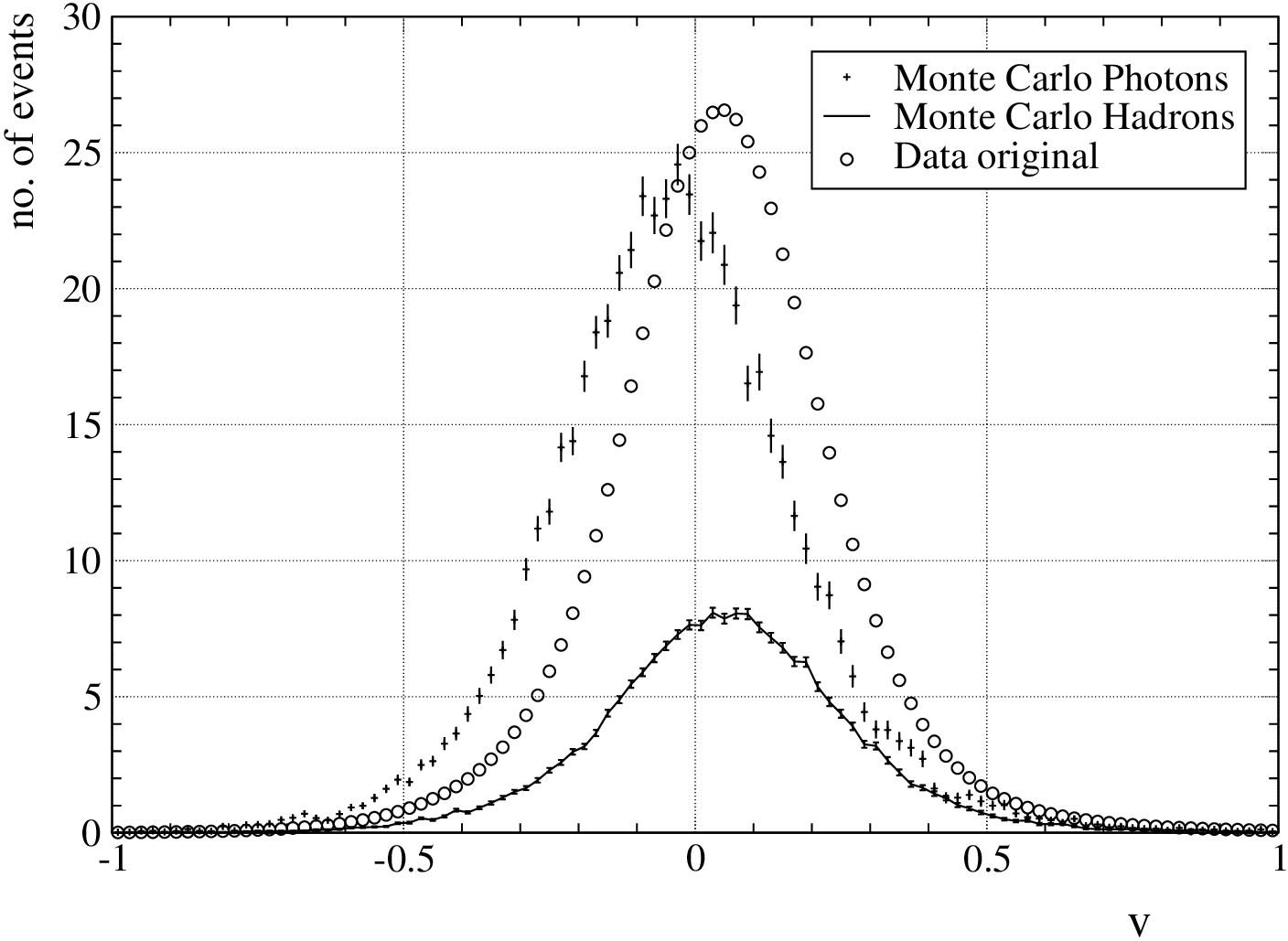} 
\caption{Distribution of the variable $v$ for MC-photon (crosses)
and MC-hadron showers (solid curve). 
The number on the y-axis is the number of 
reconstructed showers, after applying the standard cuts, assuming for
photons and hadrons the same differential 
flux at 1 TeV of 0.2582 particles / (${\rm m}^2\cdot {\rm s}\cdot {\rm sr} 
\cdot{\rm TeV}$), the spectral indices
listed in Table \ref{table1} and an effective on-time of 1 s.
The circles represent the $v$ distribution of the experimental data, 
multiplied by a factor $6\cdot10^{-7}$.}
\label{fig10}
\end{figure}

The distributions $q_\gamma, q_h$ and $q_{data}$ are displayed in
Fig. \ref{fig10}. Due to the expected very small
contribution from photons $q_{data}$ and $q_h$ are very similar. However,
a closer look reveals that the RMS of $q_{data}$ (0.206) is slightly lower 
than the RMS of $q_h$ (0.224).
These larger fluctuations of the MC data could have their origin in the
shower simulation or may be due to a slight overestimation of the
measurement errors in the MC data (cf. Section 3.3.). To 
correct for this discrepancy
the experimental distribution will be smeared by distributing the content
of one bin over the same bin and neighbouring bins according to a
Gaussian. The sigma $s$ of the Gaussian is determined as third parameter 
in the fit.

\begin{figure}[p]
\includegraphics[clip,width=\bild]{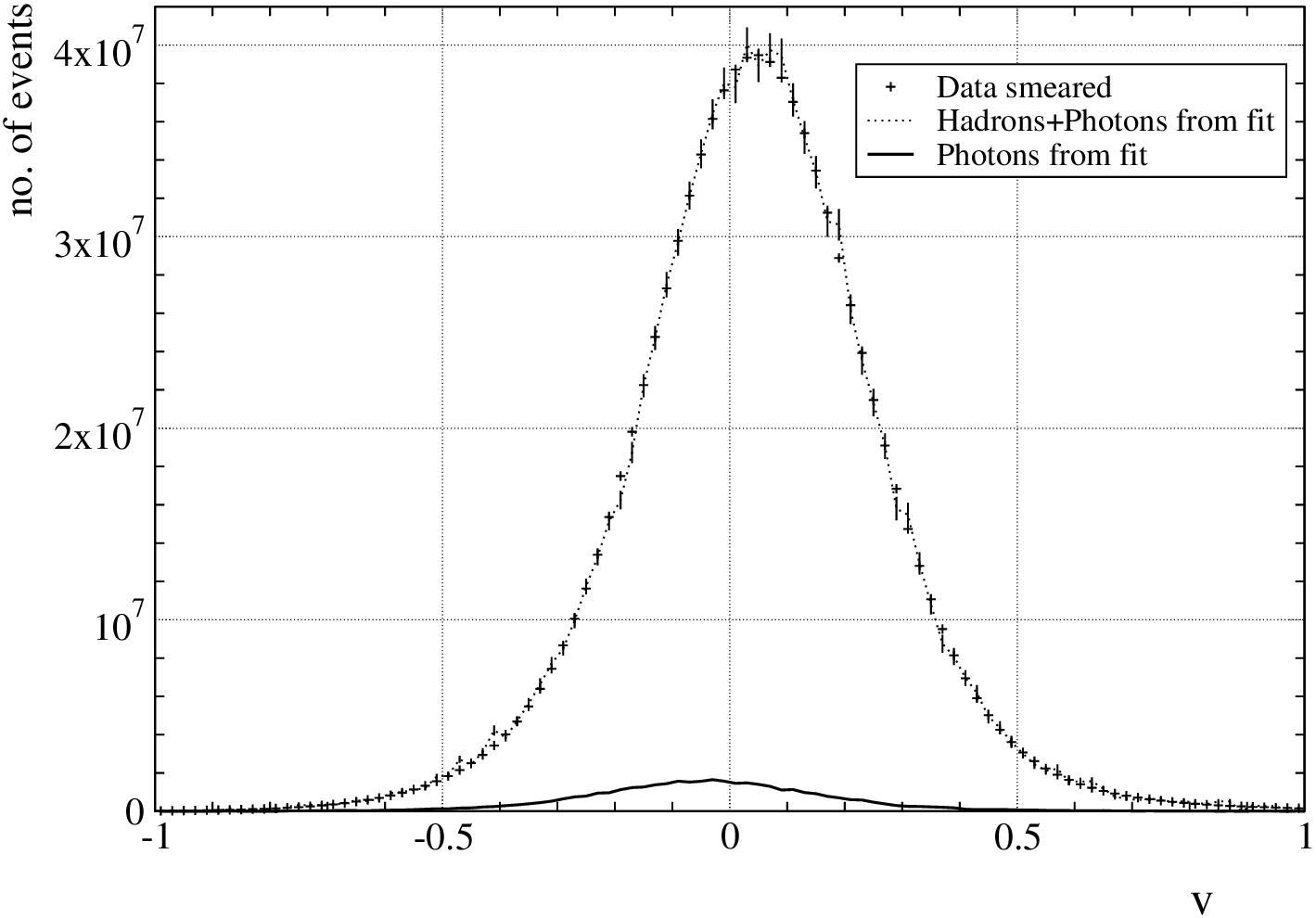} 
\caption{Distribution of the variable $v$ for the smeared experimental data 
(points with error bars). 
The dotted curve represents the fitted sum of the contributions 
from photons and hadrons, the solid curve the fitted contribution from
photons only.}
\label{fig14}
\end{figure}

The experimental distribution $q_{data}$ can be well fitted by a
superposition of $q_\gamma$ and $q_h$: the $\chi^{2}$ is 88 for 85
degrees of freedom (Fit A). In Fig. \ref{fig14} the experimental 
distribution (points with error bars) is compared with the fitted sum of 
the contributions from hadrons and photons (dotted curve). The fitted 
contribution from photons is drawn as solid curve.
The distribution of the pulls has an average
of 0.21 and an RMS of 0.99. The results for the parameters $r_0$ and $s$
are 
\vspace*{-3.0ex}
%\begin{eqnarray}
%\left. 
% \begin{array}{lll}
% r_0 & = & 0.014    \pm 0.005  \\
% s   & = & 0.080    \pm 0.003  
% \end{array}
%\right.
%\end{eqnarray} 
\begin{eqnarray}
 r_0 \;=\; 0.014    \pm 0.005\;\;\;\;\;\;\;\;\;\;\;\;
 s   \;=\; 0.080    \pm 0.003  
\end{eqnarray} 

The quoted errors correspond to 1.282 sigma, so that $r_0$ + $\Delta r_0$
is the upper limit of $r_0$ at the 90\% confidence level, taking only the
statistical errors into account. 

To check the stability of the results and to assess the systematics,
$a$ and $r_0$ were also determined as a
function of $v_0$, where $v_0$ is the border line between 2 intervals of
$v$ : $v<v_0$ and $v>v_0$. The integrated distribution of $q_{data}$,
$q_h$ and $q_\gamma$ in the 2 intervals were used to calculate $a$ and
$r_0$ (solving two equations with two unknowns), keeping $s$ fixed at 0.080. 
Both $a$ and $r_0$ are found to
be stable (and consistent with the results from Fit A)
in the region $- 0.18 < v_0 < 0.30$. This is the region which
contains the bulk of the showers: 74 \% of $q_{data}$, 76 \% of $q_h$
and 68 \% of $q_\gamma$. For $v_0$ outside of the above interval $a$ and
$r_0$ are not stable indicating that the tails of the $q_{data}$
distribution are not well represented by the MC simulations.

Changing $v_0$ from $-0.18$ to $-0.5$ results in a change of $r_0$ from
1.4\% to $-1.5$\%. Setting the parameter $s$ to zero in Fit A yields
$r_0 = -3.1\%$ with an unacceptable $\chi^2$ of 465 (for 86 degrees of
freedom). Using different variables ($(-1/R_L), \log_{10}(N_s/L_{90})$)
instead of $v$ gives either unacceptable fits ($\chi^2$ / 
(no.of degrees of freedom) $>$ 2) or $r_0$ values between 
$-3$\% and 0\%. These results suggest
that the $r_0$ from Fit A can be regarded as an upper limit with respect
to systematic errors. They also imply that $r_0$ from Fit A should not
be interpreted as a positive photon signal, although $r_0$ differs from
zero by 3.6 standard deviations. If $r_0$ is fixed at zero in Fit A the 
$\chi^2$ value increases from 88 to 101, corresponding to a change of the
$\chi^2$ probability from 38 \% to 11 \%, which is still acceptable.
The 90\% confidence upper limit of $r_0$ from Fit A is thus $r_{0,upl} 
\;=\;r_0 + \Delta r_0\;=\;1.9\times 10^{-2}$.

The ratio $r_{0,upl}$ at 1 TeV is transformed to the average photon energy
of the selected data sample of 53 TeV, using the spectral
indices of the photon and hadron fluxes adopted in the MC simulation (see
Table \ref{table1}). One obtains 
\vspace*{-3.0ex}
%\begin{tabular}{c}
%$\dfrac{I_\gamma} {I_{CR}}\; (E = 53\; \rm TeV)\;<\;1.4\times 10^{-2}\;\;\;\;\;$ 
%at 90\%\;{\rm c.l.}
%\end{tabular}
\begin{eqnarray}
\dfrac{I_\gamma} {I_{CR}}\; (\langle E\rangle = 53\; \rm TeV)\;<\;1.4\times 10^{-2}\;\;\;\;\; 
{\rm at}\;90\%\;{\rm c.l.}
\end{eqnarray} 
The results presented so far were obtained with event samples defined by the
standard selections. Another fit was performed using tighter cuts in
$\Theta$ and $(-1/R_L)$ : {$\Theta\;<\;15^\circ$ and 
$-0.01\; \rm{m}^{-1}\;<\;(-1/R_L) < 0.01\; \rm{m}^{-1}$}. 

The motivation for these selections is a stronger suppression of
very-high-energy showers ($\Theta$ cut) and a rejection of showers with
large penetration depths ( $(-1/R_L)$ cut). The latter
cut preferentially rejects hadron showers because they exhibit
larger fluctuations
in the shower development. An excellent fit is obtained with $\chi^2 = 46.8$
for 56 degrees of freedom. $r_0$ and $r_{0,upl}$ are determined as 0.012 
and 0.016 respectively, yielding 
\vspace*{-3.0ex}
%\begin{tabular}{ll}
%$\dfrac{I_\gamma} {I_{CR}}\; (E = 31\; {\rm TeV})$ &$<\;1.2\times 10^{-2} \;\;\;\;\;$ 
%at 90\%\;{\rm c.l.} 
%\end{tabular}
\begin{eqnarray}
\dfrac{I_\gamma} {I_{CR}}\; (\langle E\rangle = 31\; {\rm TeV})\;<\;1.2\times 10^{-2} \;\;\;\;\; 
{\rm at}\;90\%\;{\rm c.l.} 
\end{eqnarray} 
\subsection{Upper limit of $I_\gamma / I_{CR}$ for the inner galaxy}
The approach described in the previous section is subject to non-negligible
systematic
effects due to inconsistencies between the MC simulations and the experimental
data. One may bypass these problems by looking for a photon signal using
experimental data only. The MC simulation is then only needed to parametrize
position and shape of the expected photon contribution and to convert the
photon signal (given in terms of numbers of events) into a photon flux.

Since the diffuse gamma radiation of galactic origin is expected to be 
concentrated at the galactic disc, with a width in the galactic latitude $b$
of a few degrees, it is reasonable to compare the $v$ distribution at low
$|b|$ (signal region) with that at higher $|b|$ (background region). A
larger contribution of photons in the signal region should then show up as
an enhancement in the distribution for the signal region relative to the 
distribution for the background region. 

The acceptance of the HEGRA array depends strongly on the zenith angle
$\Theta$ (due to the construction of the detectors and
due to the varying overburden in the atmosphere) and also on the
azimuthal angle $\phi$ (because the detector plane is not horizontal). In
addition, because of the varying atmospheric conditions and because
of hardware changes of
the detector, the acceptance is a function of the time $t$.

These are the reasons why a photon signal is not searched by comparing the
$v$ distributions in absolute scale but rather by comparing their shapes.
It was found, however,  that the shapes of the experimental $v$ distributions
depend on $\Theta$ and $\phi$ too. In principle this could be due to a
non-isotropic photon radiation. However, the variations
are very likely to a large extent caused
by the same reasons which are also responsible for the systematic variations 
of the acceptance with $\Theta, \phi$ and $t$. Therefore, before comparing 
$v$ distributions from different sky regions, which obviously are taken in 
different regions of the $(\Theta,\phi,t)$ space,
one has to define the two 
$v$ distributions such that they contain the same systematic effects. This is
done in the following way :

The experimental data are divided into 12 bins in $\cos\Theta$, 18 bins in
$\phi$, 352 bins in $t$ and 22 bins in $v$. One time bin was defined as the
minimum number of full nights, with a total observation time of at least 5
hours, corresponding to a right ascension scan of $\geq 75^\circ$. 
Within each bin of
$(\cos\Theta,\phi,t)$ the $v$ distributions of photons and hadrons are assumed
to be independent of $\cos\Theta, \phi$ and $t$.

With i denoting the i-th bin in the $(\cos\Theta,\phi,t)$ space 
and j the j-th bin of $v$
the following quantities are defined : \\

\begin{supertabular}{ll}
$n_{ij}$     &no.of events from the signal region falling into the i-th bin \\ 
             &of $(\cos\Theta,\phi,t)$ and into the j-th bin of $v$ \\

$N_i = \sum_j n_{ij}$
             &total no.of signal events in the i-th bin of 
              $(\cos\Theta,\phi,t)$ \\

$v_{ij} = n_{ij} / N_i$
             &normalized $v$ distribution of events from the signal region \\ 
             &in the i-th bin of $(\cos\Theta,\phi,t)$ \\

$m_{ij}$     &as $n_{ij}$ but for events from the background region \\

$M_i = \sum_j m_{ij}$
             &total no.of background events in the i-th bin of 
              $(\cos\Theta,\phi,t)$ \\

$w_{ij} = m_{ij} / M_i$
             &normalized $v$ distribution of events from the background \\ 
             &region in the i-th bin of $(\cos\Theta,\phi,t)$ 
\end{supertabular}

Because the $v$ distribution in general depends on i, but not on $\cos\Theta$,
$\phi$ and $t$ within each bin of $(\cos\Theta,\phi,t)$, 
information about different photon contributions
in the signal and the background region could be obtained by comparing
the $v$ distributions at fixed i. One may also compare $v$ distributions
summed over all i, provided the statistical weights of the $v$ distributions
in each bin i are chosen to be the same for the signal and the background
region :

The $v$ distribution $S_j$ and its error $\Delta S_j$ for the signal region
are defined as
\vspace*{-3.0ex}
%\begin{eqnarray}
%  S_j            & = & \sum_i N_i \cdot v_{ij}  = \sum_i n_{ij} \\
%  (\Delta S_j)^2 & = & \sum_i n_{ij}                 
%\end{eqnarray} 
\begin{eqnarray}
  S_j           =  \sum_i N_i \cdot v_{ij}  = \sum_i n_{ij}\;\;\;\;\;\;\;\;\;\;
  (\Delta S_j)^2=  \sum_i n_{ij}                 
\end{eqnarray} 
The $v$ distribution $B_j$ and its error $\Delta B_j$ for the background
region are now constructed as
\vspace*{-3.0ex} 
\begin{eqnarray}
   B_j           =  \sum_i N_i \cdot w_{ij} \;\;\;\;\;\;\;\;\;\;\;\;\;\;\;\;\;
\;\;\;\;\;\;   (\Delta B_j)^2=  \sum_i N^2_i \cdot (\Delta w_{ij})^2 \\
\;\;\;\;\;\;\;\;\;\;\;\;\;{\rm with}\;\;(\Delta w_{ij})^2\;=\; 
                                m_{ij}\cdot(M_i - m_{ij}) / M^3_i \nonumber
\end{eqnarray} 
By weighting the $v$ distribution $w_{ij}$ 
with $N_i$ the systematic effects which are
present in $S_j$ are simulated in $B_j$. Differences between the shapes 
of the $S_j$ and $B_j$ distributions are now interpreted as being due to 
different contributions of photons. 

This approach is similar to the standard approach proposed by 
\cite{alexandreas}.  
However, in contrast to \cite{alexandreas}, in the present analysis

\begin{itemize}
\item[a)] the background distribution $w_{ij}$ is not determined from all
events but only from those in a well defined sky region, which is different 
from and has no overlap with the signal region. Therefore signal and
background distributions ($S_j$ and $B_j$ respectively) are statistically
independent. In addition, the resulting upper limit on the photon
contribution is more realistic (in general higher) as compared to the case
where signal and background regions do overlap. In the latter case the photon
contribution would be underestimated in general. \\

\item[b)] the background distribution $B_j$ is not constructed by randomly
generating a $v$ value from the $w_{ij}$ distribution but rather by using
for each observed event the full $w_{ij}$ distribution. In this way full use
is made of the measured $w_{ij}$ distribution.
Both a) and b) make the error calculation simple and transparent. \\

\item[c)] the sums $\sum_i$ in the calculation of $S_j,\;B_j$ and their errors
extend only over those i for which $M_i$ is greater than $N_i$. This means that
in $S_j$ and $B_j$ only those $(\cos\Theta,\phi,t)$ bins are considered for
which the background distribution $w_{ij}$ is at least as well measured
as the signal distribution $v_{ij}$. In particular, if 
there are no background events in a certain $(\cos\Theta,\phi,t)$ bin 
(i.e. $M_i\;=\;0$) then the corresponding events in the signal region 
($N_i$) are not included in the calculation of $S_j$. The requirement
$M_i\;>\;N_i$ thus removes possible biases in the comparison of $S_j$ and
$B_j$. Of course, one could think of less restrictive conditions on the
bins to be included in the sums, for example demanding $M_i$ to be greater 
than some positive value, independent of $N_i$.
\end{itemize}

The ratio $r_0 = I^0_\gamma / I^0_{CR}$ at 1 TeV is determined by fitting
the signal distribution $S_j$ by a superposition of the background distribution
$B_j$ and a term $G_j$, representing a possible contribution from photons :
\vspace*{-3.0ex}
\begin{eqnarray}
S_j &= &a\cdot\left( B_j + r_0\cdot \dfrac{G_j}{\epsilon} \right)\;\;\;\;\;
       (j=1,...22)
\end{eqnarray}
For $G_j$ the MC expectation $q_\gamma$ (see above) for reconstructed photon 
showers, rebinned and normalized to $\sum_i N_i$ 
reconstructed events, is taken. $\epsilon$ is an average ratio of the
number of reconstructed hadron to the number of reconstructed photon
showers,
assuming the same differential flux of photons and hadrons at 1 TeV. The
parameters $a$ and $r_0$ are adjusted in the fit.

The signal and background regions are defined as : 

\begin{tabular}{ll}
 \hspace*{0.5cm}signal region :     &$20^\circ\;<\;l\;<\;60^\circ,$ and 
 $|b|\;<\;5^\circ$  \\
 \hspace*{0.5cm}background region : &$20^\circ\;<\;l\;<\;60^\circ,$ and 
 $10^\circ <|b|\;<\;30^\circ $  
\end{tabular} 

\begin{figure}[p]
\includegraphics[clip,width=\bild]{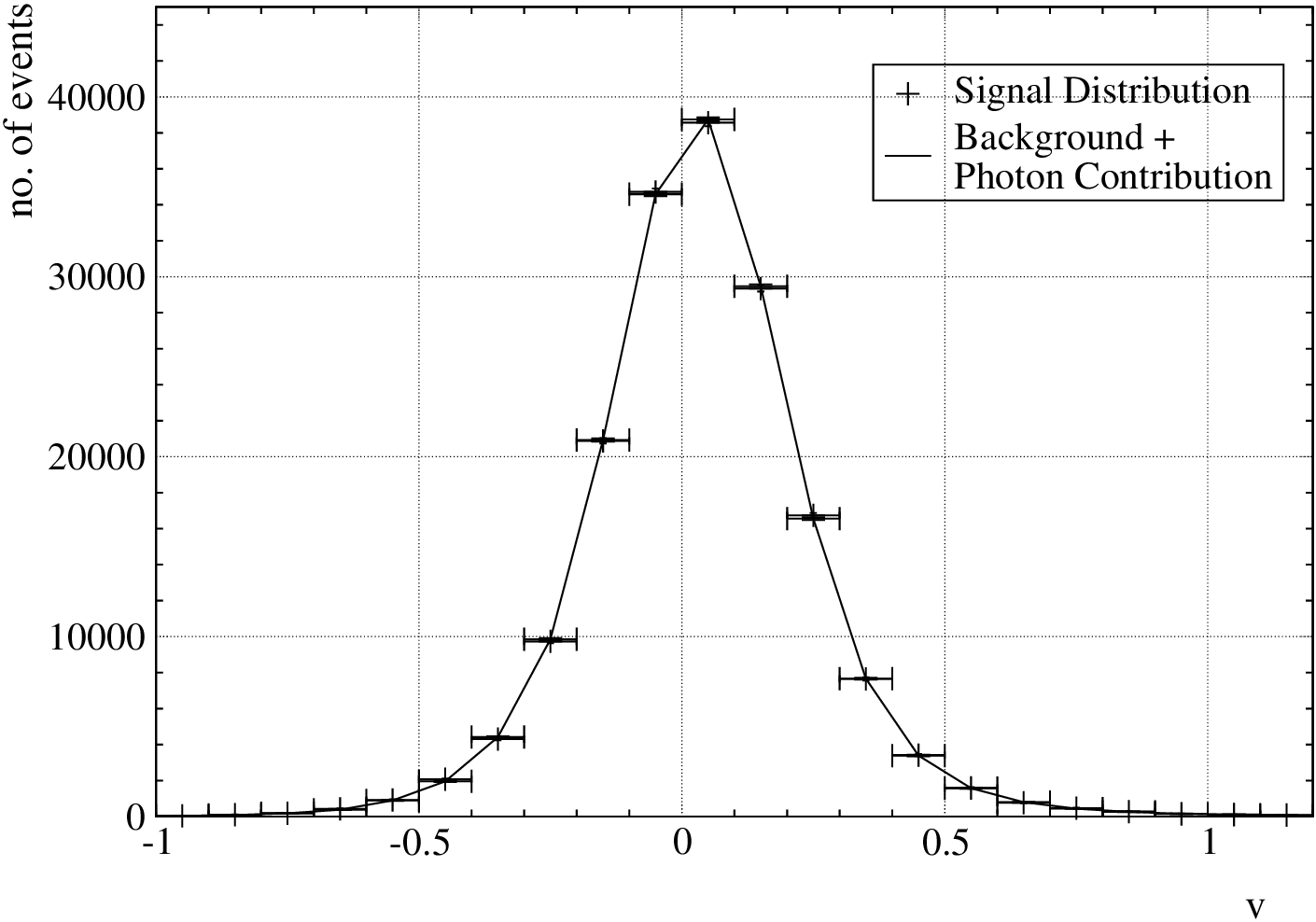} 
\caption{Distribution $S_j$ of the variable $v$ for the signal region 
(data points with horizontal bars). 
The polygone represents the fitted sum of the contributions 
from the background region ($a\cdot B_j$) and from photons
$(a\cdot r_0 \cdot \dfrac{G_j}{\epsilon})$.}
\label{figA}
\end{figure}

With this definition the average $l$ of the events in the signal region 
is $48.6^\circ$ and $\epsilon$ is calculated as 0.348. The average $l$ and
$|b|$ of the events in the background region is $45.7^\circ$ and $20^\circ$
respectively. The results of the fit are : 
\vspace*{-3.0ex}
\begin{eqnarray}
 a       \;=\;   1.0007  \pm 0.0094\;\;\;\;\;\;\;\;\;\;\;\;
 r_0     \;=\; -(0.0003  \pm 0.0030) 
\end{eqnarray} 
with a $\chi^2$ of 26.6 for 20 degrees of freedom. The signal distribution
$S_j$ is compared with the fitted sum of the 
background distribution $(a\cdot B_j)$
and the contribution from photons $(a\cdot r_0\cdot\dfrac{G_j}{\epsilon})$
in Fig. \ref{figA}. 

\begin{figure}[p]
\includegraphics[clip,width=\bild]{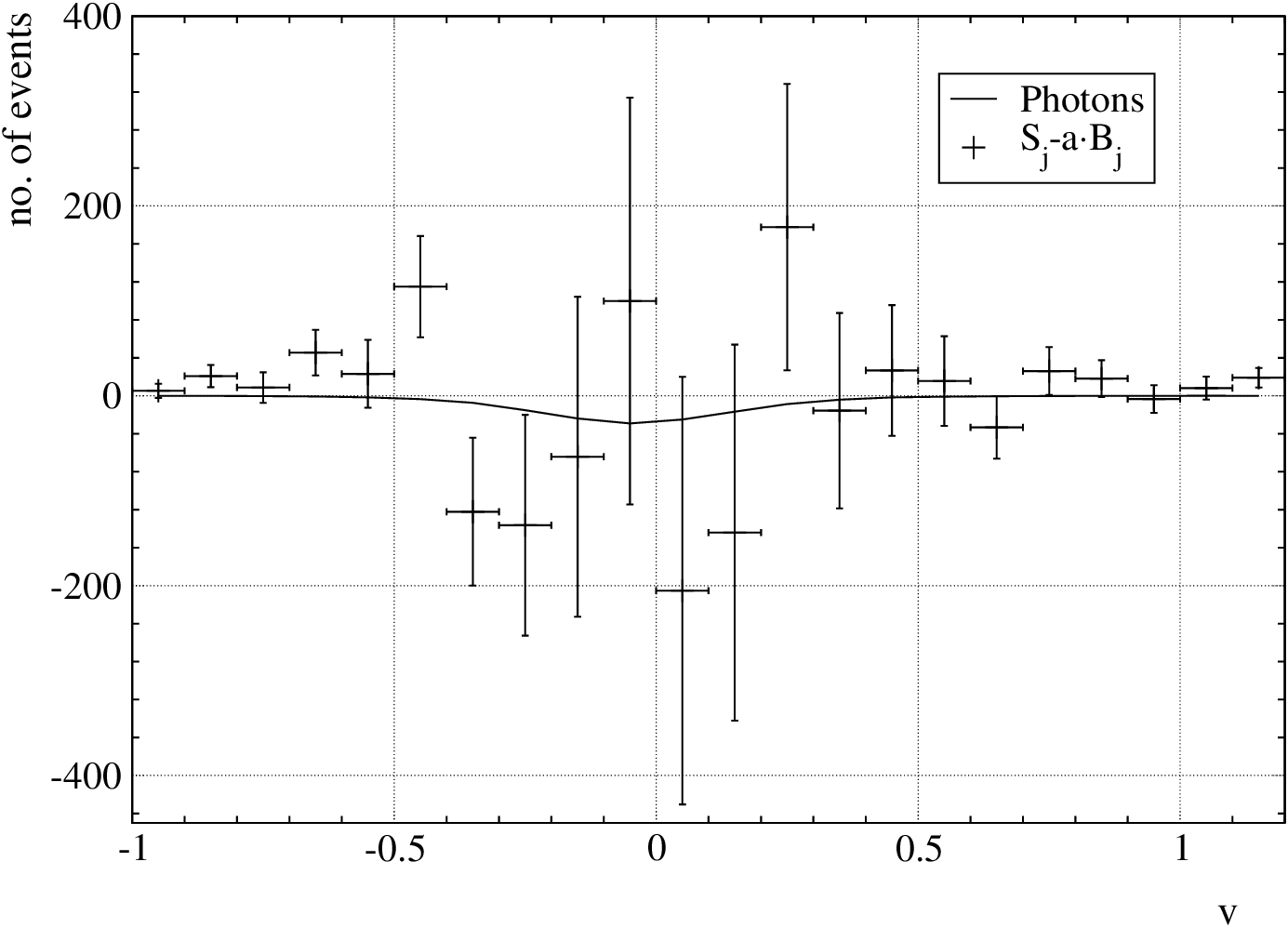} 
\caption{Difference $(S_j - a\cdot B_j)$ between the $v$ distributions of the
signal and the background region (points with error bars). 
The solid curve represents the fitted 
contribution from photons $(a\cdot r_0 \cdot \dfrac{G_j}{\epsilon})$.}
\label{figB}
\end{figure}

Transforming $r_0$ to the average energy of reconstructed photon showers
from the signal region of 54 TeV gives the upper limit 
\vspace*{-3.0ex}
%\begin{tabular}{c}
%$\dfrac{I_\gamma} {I_{CR}}\; (E = 54\; \rm TeV) < 2.0\cdot 10^{-3}\;\;\;\;\;$ 
%at 90\%\;{\rm c.l.} \\
%\end{tabular}
\begin{eqnarray}
\dfrac{I_\gamma} {I_{CR}}\; (\langle E\rangle = 54\; \rm TeV)\;<\;2.0\times 10^{-3}\;\;\;\;\; 
{\rm at}\;90\%\;{\rm c.l.} 
\end{eqnarray} 
The difference $(S_j - a\cdot B_j)$ is shown in Fig. \ref{figB}, together with 
the fitted
contribution from photons $(a\cdot r_0\cdot\dfrac{G_j}{\epsilon})$.  
The distribution of the pull values has a mean of $-0.43$ and an RMS of 0.99. 

\section{Discussion of the Results}
It should be noted that in the present analysis there is no way of 
discriminating photon-induced showers from showers induced by electrons or
positrons. Therefore the quoted upper limits are limits for the sum of the
contributions from photons, electrons and positrons. Extrapolating the 
differential electron flux (using a spectral index of 3.3), 
as measured by \cite{nishimura} between 30 GeV and 1 TeV, 
and comparing with the predicted
flux of diffuse photons at 50 TeV \cite{porter1} yields an $e^-/\gamma$
ratio in the order of 10\%, at 50 TeV for the inner galaxy. The contribution
from positrons is probably an order of magnitude lower than that from
electrons \cite{musser}. The measured upper limits are therefore in good 
approximation upper limits of the diffuse photon flux.

\begin{figure*}[p]
\begin{center}
\includegraphics[clip,width=12cm]{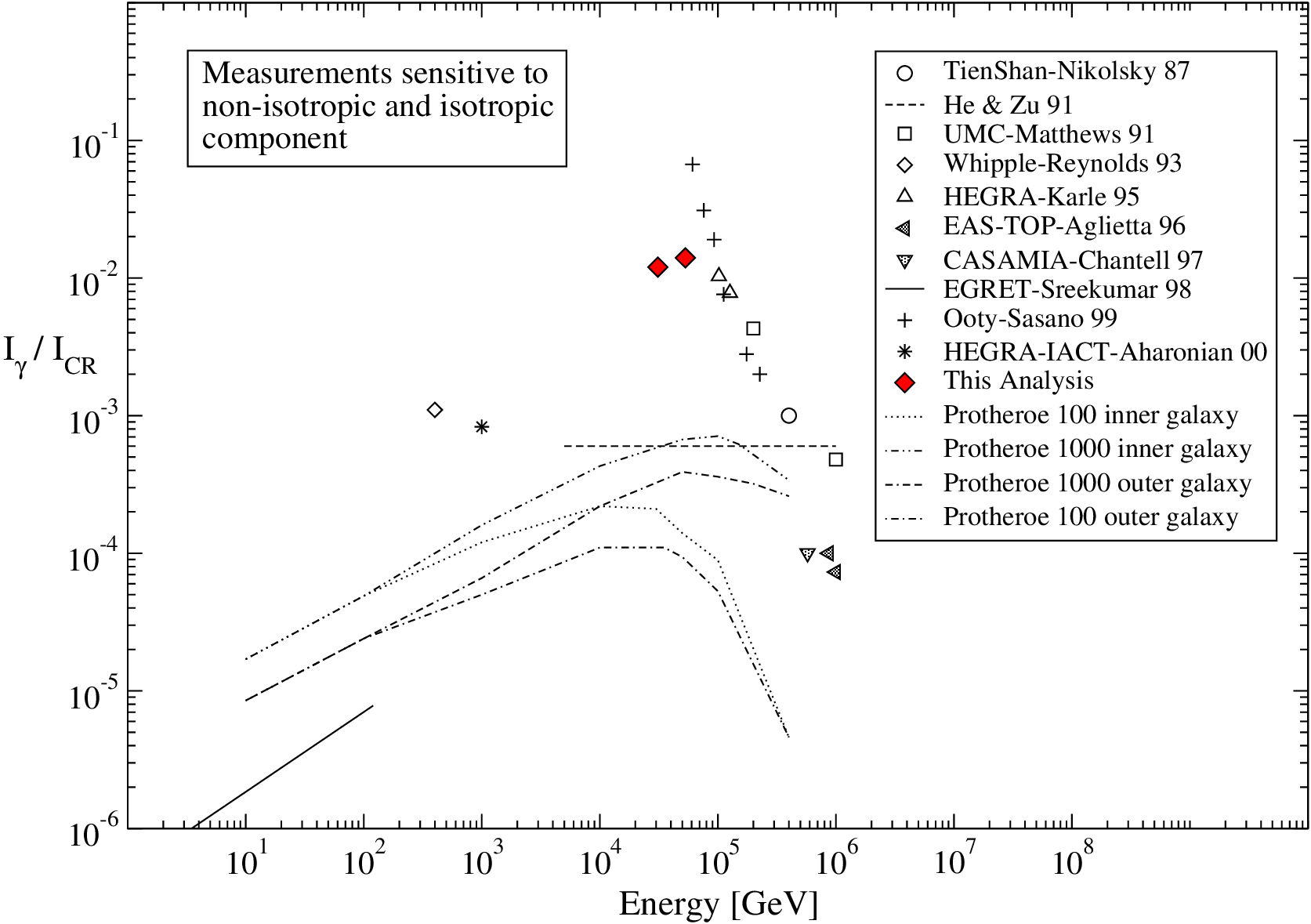} 
\end{center}
\caption[cap15]{Comparison of measurements of the ratio $I_\gamma / I_{CR}$ 
as a
function of the energy in the 10 GeV to 1 PeV range. All data points are upper 
limits except those from Tien-Shan
(\cite{nikolsky}) and from EGRET (\cite{sreekumar}). The solid line is the
EGRET measurement of the isotropic component only. Upper limits 
are plotted only from those measurements which are sensitive to both the
non-isotropic and the isotropic component of the diffuse photon background.
The curves represent theoretical predictions by \cite{porter1} with the 
following specifications : cutoff energy of the
electron injection spectrum 100 TeV or 1000 TeV respectively, inner galaxy
($-60^{\circ} < l < 60^{\circ}, |b| < 10 ^{\circ}$) and outer galaxy
($50^{\circ} < l < 220^{\circ}, |b| < 10 ^{\circ}$) respectively. For all
predictions the spectral index of the electron injection spectrum was assumed
to be equal to 2.0 .}
\label{fig15}
\end{figure*}

\begin{figure*}[p]
\begin{center}
\includegraphics[clip,width=12cm]{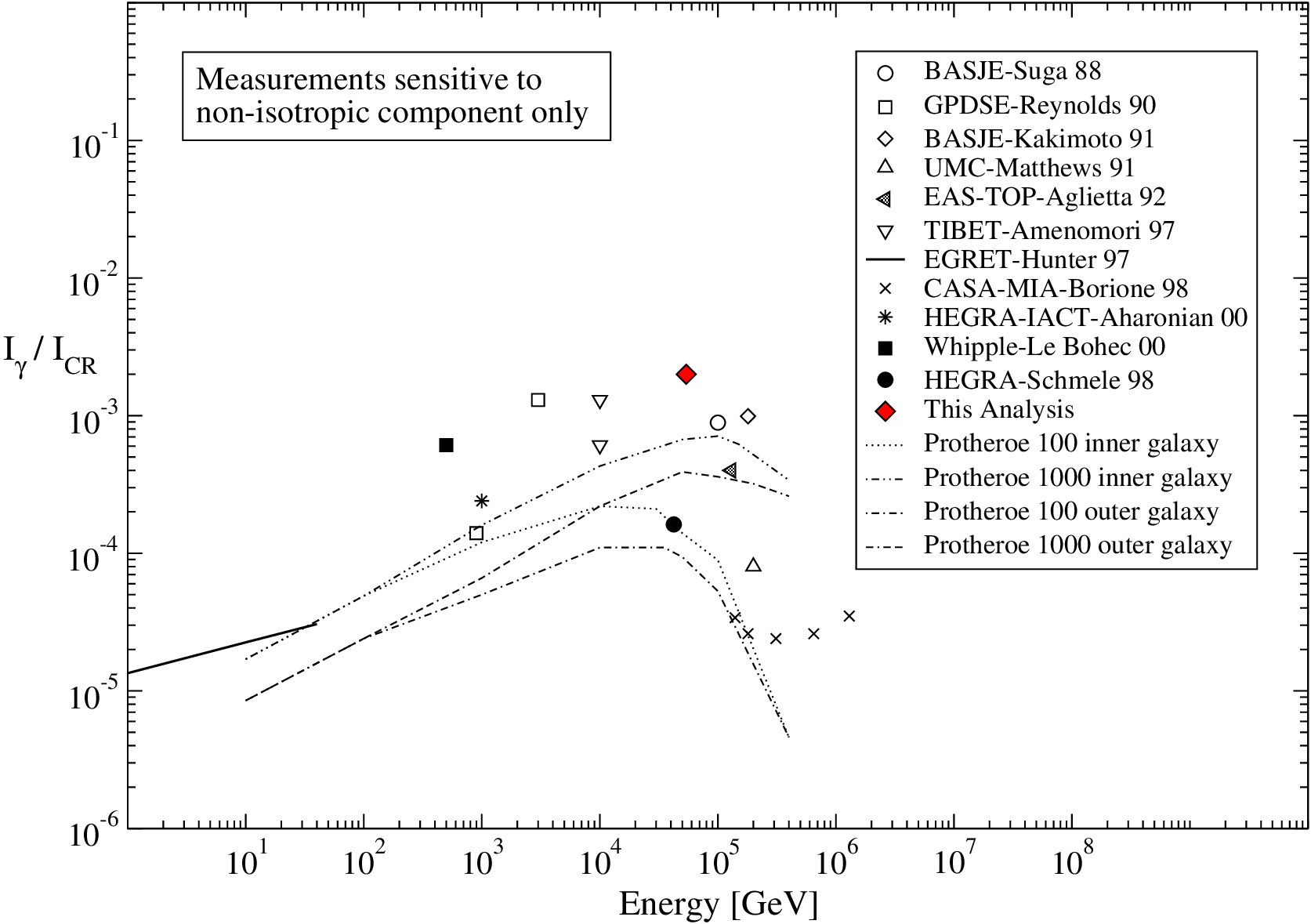} 
\end{center}
\caption[cap16]{Comparison of measurements of the ratio $I_\gamma / I_{CR}$ as a
function of the energy in the 10 GeV to 1 PeV range. All data points are upper 
limits except those from BASJE
(\cite{suga}) and from EGRET (\cite{hunter}). The solid line is
the EGRET measurement of the non-isotropic component only. Upper limits 
are plotted only from those measurements which are sensitive to the
non-isotropic component of the diffuse photon background only.
The curves represent theoretical predictions by \cite{porter1} with the 
following specifications : cutoff energy of the
electron injection spectrum 100 TeV or 1000 TeV respectively, inner galaxy
($-60^{\circ} < l < 60^{\circ}, |b| < 10 ^{\circ}$) and outer galaxy
($50^{\circ} < l < 220^{\circ}, |b| < 10 ^{\circ}$) respectively. For all
predictions the spectral index of the electron injection spectrum was assumed
to be equal to 2.0 .}
\label{fig16}
\end{figure*}

The existing measurements of $I_\gamma / I_{CR}$ in the energy region
from 10 GeV to 1 PeV are displayed in Figs. \ref{fig15} and \ref{fig16} 
and compiled in Tables \ref{table2} and \ref{table5}. 
Fig. \ref{fig15} displays the measurements
which are sensitive to both the isotropic and the non-isotropic component of 
the diffuse photon flux. The measurements which are sensitive only to the 
non-isotropic component are plotted in Fig. \ref{fig16}.

With one exception \cite{he}, 
at fixed energy all upper limits in Fig. \ref{fig15}
(Table \ref{table2}) are bigger than those in Fig. \ref{fig16} (Table
\ref{table5}). This reflects the larger systematic errors in measurements 
which are based on absolute comparisons of experimental data with MC 
predictions. In the experiment \cite{he}
the flux and the energy distribution of secondary photons and hadrons is
measured at various depths in the atmosphere. It uses the different
attenuation lengths for photon-induced showers and
for the electromagnetic component of 
hadron-induced showers to determine an upper limit of $I_\gamma / I_{CR}$
in the energy region 5 TeV to 1 PeV. In \cite{karle1} the experiment is 
critisized by claiming that the theoretical errors due to uncertainties in the
hadronic interaction model and in the chemical composition may be
underestimated.

Compared to the present analysis the first measurement of a diffuse photon 
flux obtained from HEGRA-array data \cite{karle1} 
was based on a much smaller statistics of experimental data
(by a factor of 250) and of MC data (by a factor of 55). In addition the
detector consisted of 169 scintillation detectors (now 182) and 49 
Cherenkov detectors (now 97). Nevertheless the old result for $r$ is
comparable to the new result, which refers to slightly lower energies. In
the present analysis more emphasis is put on a good MC tuning. 
Unfortunately it turns out that because of the increased fluctuations in
the MC data (see Sect.3.3) and
because of the lower energies, which imply
larger errors in the measurements, the potential for a photon-hadron
discrimination is clearly reduced.

The upper limits in Fig. \ref{fig15} (Table \ref{table2}), including the 
result of the present analysis, are in general larger than the theoretical
expectations by 1 to 2 orders of magnitude.

The measurements compiled in Fig. \ref{fig16} (Table \ref{table5}) are
insensitive to the isotropic component of the diffuse photon flux because 
they are obtained by comparing experimental distributions in a certain sky
region (usually low absolute galactic latitudes) with those in a background 
region (usually larger absolute galactic latitudes).

The measurements provide a good upper limit on the non-isotropic
component of the diffuse photon flux only if this component is negligible in
the background region. Otherwise the upper limits will in general
be too low. In the
present analysis the background region is well separated from the signal
region ($5^\circ$ in galactic latitude) in order to fullfill this condition.
Note that if the IC process is dominating \cite{porter,dar,strong1} a rather 
broad distribution of the diffuse photon flux in the
galactic latitude is expected, in which case a good separation between signal
and background region is even more critical.

The upper limit from this analysis is still compatible with the theoretical
predictions shown in Fig. \ref{fig16}. 
The measurements from \cite{matthews,schmele,borione} on the other hand 
suggest that in the
model by \cite{porter1} the cutoff energy of the electron injection spectrum 
is well below 1000 TeV. 

The result from \cite{schmele} is another HEGRA measurement in which only data
from the scintillation detectors were used. In that analysis the lack of
any gamma-hadron discrimination is more than compensated by the very
large data sample, collected during 2 2/3 years of data taking 
(day and night). In contrast to the data from the Cherenkov detectors,
the data from the scintillation detectors depend on the atmospheric
conditions (pressure and temperature) in a well controllable way and
are practically insensitive to other conditions like the humidity of the air
and the presence of clouds.

It should be emphasized that the measurements 
compiled in Table \ref{table5} and plotted in Fig. \ref{fig16} do not
refer to the same sky regions and are therefore strictly speaking not
directly comparable. This should also be kept in mind when the experimental 
upper limits are compared with the theoretical predictions, which are for 
very specific sky regions.

\section*{Acknowledgements}
 The support of the HEGRA experiment by the German Ministry for Research 
and Technology BMBF and by the Spanish Research Council
CYCIT is acknowledged. We are grateful to the Instituto
de Astrofisica de Canarias for the use of the site and
for providing excellent working conditions on La Palma. We gratefully
acknowledge the technical support staff of Heidelberg,
Kiel, Munich and Yerevan.

\begin{table*}[p]
\begin{center}
\caption{Absolute differential fluxes $\phi = \phi_0\cdot(E/\rm TeV)^{-\gamma}$
used for the definition of MC samples.}
\label{table1}
\vspace{0.3cm}
\begin{tabular}{|l|c|c|l|c|}
\hline
generated particle & $\phi_0$ & spectral index $\gamma$ &  sample & Z  \\
\rule[-3ex]{0cm}{3ex}
 & $\left[\dfrac{\rm no.\; of\; particles}{\rm m^{2}\cdot s\cdot sr\cdot
 \rm\; TeV}\right]$   &           &  is representative for  & \\             
\hline
H           & 0.1091    &  2.75    & H         & 1 \\
He          & 0.06808   &  2.62    & He - Li   & 2 - 3 \\
0           & 0.05182   &  2.65    & Be - Si   & 4 - 14 \\
Fe          & 0.02919   &  2.60    & P - Ni    & 15 - 28 \\
&&&&  \\
H - Fe      & 0.2582    &   --     & hadrons   & 1 - 28 \\
&&&&  \\
$\gamma$    & 0.0002582 &  2.75    &  photons, $e^+, e^-$  & -- \\
\hline
\end{tabular}
\end{center}
\end{table*}

%------------------------------------ 
\begin{table*}[p]
\caption{Measurements of $I_\gamma / I_{CR}$ in the 10 GeV to 1 PeV energy
range which are sensitive to both the isotropic and the non-isotropic
component of the diffuse photon flux. 
For transforming absolute fluxes into relative fluxes a cosmic ray flux
of $I_{CR} = 0.2582 \times (E/{\rm TeV})^{-2.68} 
(\rm{m}^2 \cdot \rm{s} \cdot \rm{sr} \cdot \rm{TeV})^{-1}$ has been 
assumed. All data points are upper limits except those from Tien-Shan
(\cite{nikolsky}) and from EGRET \cite{sreekumar}. The EGRET results are those
for the isotropic component only. $l$ and $b$ denote the 
galactic longitude and latitude respectively. 
Abbreviations: ASA = air-shower array, IACT = imaging air Cherenkov telescope,
GPl = Galactic plane.  }

\begin{center}   
\label{table2}
\vspace{0.3cm}
\begin{tabular}{|l|c|c|c|l|}
\hline
experiment & references & $E_\gamma$ & $I_\gamma / I_{CR}$ & special
selections \\
\hline
&&&&  \\
Tien-Shan-ASA  & \cite{nikolsky}& $ > 400\; \rm TeV$ & $(1.0 \pm 0.3)
                 \times 10^{-3}$ & $b > 50^\circ$  \\

emulsion    & \cite{he}      & 5 TeV - 1 PeV  & $ < 6\times 10^{-4}$ & \\  

UMC-ASA     & \cite{matthews}& $ > 200\; \rm TeV$ 
                             & $ < 4.3\times 10^{-3}$                &  \\ 
            &      & $ > 1000\; \rm TeV$ & $ < 4.8\times 10^{-4}$    &  \\

Whipple-IACT & \cite{reynolds1}& $ > 400\; \rm GeV$ & $ < 1.1\times 10^{-3}$
             &  \\

HEGRA-ASA    & \cite{karle1} & 65 - 160 TeV   & $ < 1.0\times 10^{-2}$
             &  \\
             &               & 80 - 200 TeV   & $ < 7.8\times 10^{-3}$
             &  \\

EAS-TOP-ASA  & \cite{aglietta1}& $ > 1000\; \rm TeV$ & $ < 7.3\times 10^{-5}$
             &  \\
             &               & $ > 870\; \rm TeV$  & $ < 1\times 10^{-4}$
             &  \\

CASA-MIA     & \cite{chantell}& $ > 575\; \rm TeV$  & $ < 10^{-4}$   &  \\

EGRET        & \cite{sreekumar}& 10 GeV   & $1.8\times 10^{-6}$ 
             & out of GPl  \\
             &                & 100 GeV  & $7.0\times 10^{-6}$
             & out of GPl  \\

Ooty-ASA     & \cite{sasano}  & 61 TeV   & $ < 6.7\times 10^{-2}$  &   \\
             &                & 76 TeV   & $ < 3.1\times 10^{-2}$  &   \\
             &                & 93 TeV   & $ < 1.9\times 10^{-2}$  &   \\
             &                & 112 TeV  & $ < 7.6\times 10^{-3}$  &   \\
             &                & 175 TeV  & $ < 2.8\times 10^{-3}$  &   \\
             &                & 227 TeV  & $ < 2.0\times 10^{-3}$  &   \\

HEGRA-IACT   & \cite{aharonian4}& 1 TeV & $ < 8.3\times 10^{-4}$ 
             & GPl, $37^\circ < l < 43^\circ, |b| < 5^\circ$  \\

HEGRA-ASA    & this           & 31 TeV   & $ < 1.2\times 10^{-2}$ &  \\
             & analysis       & 53 TeV   & $ < 1.4\times 10^{-2}$ &  \\
\hline
\end{tabular}
\end{center}
\end{table*}

\clearpage
%------------------------------------
\begin{table*}[p]
\caption{Measurements of $I_\gamma / I_{CR}$ in the 10 GeV to 1 PeV energy
range which are sensitive to the non-isotropic component of the diffuse 
photon flux only. 
For transforming absolute fluxes into relative fluxes a cosmic ray flux
of $I_{CR} = 0.2582 \times (E/{\rm TeV})^{-2.68} 
(\rm{m}^2 \cdot \rm{s} \cdot \rm{sr} \cdot \rm{TeV})^{-1}$ has been 
assumed. All data points are upper limits except those from BASJE
(\cite{suga}) and from EGRET \cite{hunter}. The EGRET results are those for
the non-isotropic component only. $l$ and $b$ denote the 
galactic longitude and latitude respectively. $\delta$ and $\alpha$ are 
declination and right ascension respectively.
Abbreviations: ASA = air-shower array, IACT = imaging air Cherenkov telescope,
GPDSE = galactic plane drift scan experiment, GPl = Galactic plane.  }

\begin{center}   
\label{table5}
\vspace{0.3cm}
\begin{tabular}{|l|c|c|c|l|}
\hline
experiment & references & $E_\gamma$ & $I_\gamma / I_{CR}$ & special
selections \\
\hline

BASJE-ASA      & \cite{suga}  & $ > 100\; \rm TeV$ & $8.9\times 10^{-4}$
            &  $-40^\circ < \delta < 0, 180^\circ < \alpha < 210^\circ$  \\

GPDSE       & \cite{reynolds} & $ > 900\; \rm GeV$ & $ < 1.4\times 10^{-4}$
            & GPl, $|b| < 5^\circ$  \\
            &               & $ > 3\; \rm TeV$   & $ < 1.3\times 10^{-3}$
            & GPl, $|b| < 5^\circ$  \\ 

BASJE-ASA   & \cite{kakimoto} & $ > 180\; \rm TeV$                
                             & $ < 9.9\times 10^{-4}$ 
            &  $-40^\circ < \delta < 0, 180^\circ < \alpha < 210^\circ$  \\

UMC-ASA     & \cite{matthews} & $ > 200\; \rm TeV$  & $ < 8.0\times 10^{-5}$    
            & GPl, $30^\circ < l < 220^\circ, |b| < 10^\circ$  \\

EAS-TOP-ASA & \cite{aglietta}& $ > 130\; \rm TeV$ & $ < 4\times 10^{-4}$
            & GPl, $|b| < 5^\circ$    \\

TIBET-ASA    & \cite{amenomori}& 10 TeV   & $ < 6.1\times 10^{-4}$
             & GPl, $140^\circ < l < 225^\circ, |b| < 5^\circ$  \\
             &                & 10 TeV   & $ < 1.3\times 10^{-3}$
             & GPl, $20^\circ < l < 55^\circ, |b| < 5^\circ$    \\

EGRET        & \cite{hunter}  & 17 GeV   & $2.5\times 10^{-5}$ 
             & GPl, $300^\circ < l < 60^\circ, |b| < 10^\circ$  \\
             &                & 39 GeV   & $3.0\times 10^{-5}$
             & GPl, $300^\circ < l < 60^\circ, |b| < 10^\circ$  \\

HEGRA-ASA    & \cite{schmele,horns}& $> 42\; \rm TeV$ & $ < 1.6\times 10^{-4}$
             & GPl, $  0^\circ < l < 255^\circ, |b| < 5^\circ$ \\

CASA-MIA     & \cite{borione} & 140 TeV  & $ < 3.4\times 10^{-5}$ 
             & GPl, $50^\circ < l < 200^\circ, |b| < 5^\circ$   \\

             &                & 180 TeV  & $ < 2.6\times 10^{-5}$ 
             & GPl, $50^\circ < l < 200^\circ, |b| < 5^\circ$   \\

             &                & 310 TeV  & $ < 2.4\times 10^{-5}$ 
             & GPl, $50^\circ < l < 200^\circ, |b| < 5^\circ$   \\

             &                & 650 TeV  & $ < 2.6\times 10^{-5}$ 
             & GPl, $50^\circ < l < 200^\circ, |b| < 5^\circ$   \\

             &                & 1300 TeV  & $ < 3.5\times 10^{-5}$ 
             & GPl, $50^\circ < l < 200^\circ, |b| < 5^\circ$   \\

HEGRA-IACT   & \cite{aharonian4} & 1 TeV & $ < 2.4\times 10^{-4}$
             & GPl, $37^\circ < l < 43^\circ, |b| < 2^\circ$  \\

Whipple-IACT & \cite{bohec}   & $ > 500\; \rm GeV$ & $ < 6.1\times 10^{-4}$
             & GPl, $38.5^\circ < l < 41.5^\circ, |b| < 2^\circ$  \\

HEGRA-ASA    & this           & 54 TeV   & $ < 2.0\times 10^{-3}$
             & GPl, $20^\circ < l < 60^\circ, |b| < 5^\circ$ \\
             & analysis       &          &                      &  \\
\hline
\end{tabular}
\end{center}
\end{table*}

\clearpage

%------------------------------------

%***************************************************
%\clearpage

%\clearpage

%\clearpage

%\begin{figure}[p]
%\includegraphics[clip,width=\bild]{fig4.eps} 
%\caption{Distribution of $\Delta x_{core}$ for the experimental data
%(histogram) and for the MC data (open circles).}
%\label{fig4}
%\end{figure}
%\clearpage

%\begin{figure}[p]
%\includegraphics[clip,width=\bild]{fig5.eps} 
%\caption{Distribution of $\Delta y_{core}$ for the experimental data
%(histogram) and for the MC data (open circles).}
%\label{fig5}
%\end{figure}
%\clearpage

%\begin{figure}[p]
%\includegraphics[clip,width=\bild]{fig6_neu.eps} 
%\caption{Distribution of $\Delta\Theta$ for the experimental data
%(histogram) and for the MC data (open circles).}
%\label{fig6}
%\end{figure}
%\clearpage

%\begin{figure}[p]
%\includegraphics[clip,width=\bild]{fig7.eps} 
%\caption{Distribution of $\Delta \ln(L_{90})$ for the experimental data
%(histogram) and for the MC data (open circles).}
%\label{fig7}
%\end{figure} 
%\clearpage

%\clearpage

%\clearpage

%\clearpage

%\clearpage

%\begin{figure}[p]
%\includegraphics[clip,width=\bild]{fig12.eps} 
%\caption{Distribution of the variable $v$ for the experimental data.}
%\label{fig12}
%\end{figure}
%\clearpage

%\clearpage

%\begin{figure}[p]
%\includegraphics[clip,width=\bild]{fig11.eps} 
%\caption{The pull value as a function of the variable $v$. The pull value is 
%defined as  
%$(q_{fitted}-q_{data})\; / \; 
% \sqrt{ (\Delta q_{fitted})^2 + (\Delta q_{data})^2 }$.}
%\label{fig11}
%\end{figure}
%\clearpage

%\clearpage

%\clearpage

%\begin{figure}[p]
%\includegraphics[clip,width=\bild]{figC.eps} 
%\caption{The pull value as a function of $v$. The pull value is defined as
%$\left(a\cdot\left[ B_j+r_0\cdot \dfrac{G_j}{\epsilon}\right] -S_j\right)\;/\;
%\sqrt{\Delta^2(a\cdot B_j)+\Delta^2(S_j)}$.}
%\label{figC}
%\end{figure}
%\clearpage

%\clearpage

%\clearpage

\clearpage
%***************************************************


\begin{thebibliography}{9}

\bibitem{hunter} Hunter, S.D., et al., 1997, AJ 481, 205

\bibitem{sreekumar} Sreekumar, P., et al., 1998, AJ 494, 523

\bibitem{chiang} Chiang, J. \& Mukherjee, R., 1998, ApJ 496, 752

\bibitem{bertsch} Bertsch, D.L., et al., 1993, ApJ 416, 587

\bibitem{mori} Mori, M., 1997,
 ApJ 478, 225

\bibitem{gralewicz} Gralewicz, P., et al., 1997, 
 A\&A 318, 925

\bibitem{moskalenko98} Moskalenko, I.V., et al., 1998,
 A\&A 338, L75

\bibitem{aharonian00} Aharonian, F.A. \& Atoyan, A.M., 2000,
 A\&A 362, 937

\bibitem{porter} Porter, T.A. \& Protheroe, R.J., 1997,
 J. Phys. G23, 1765

\bibitem{pohl} Pohl, M. \& Esposito, J.A., 1998, ApJ 507, 327

\bibitem{porter1} Porter, T.A. \& Protheroe, R.J., 1999, Proc. 26th
 ICRC, Salt Lake City, OG 3.2.38

\bibitem{strong} Strong, A.W., et al., 1999, ApL 38, 445 

\bibitem{dar} Dar, A. \&  De Rujula, A., 2000, astro-ph/0005080

\bibitem{strong1} Strong, A.W., et al., 2000, ApJ 537, 763

\bibitem{moskalenko1} Moskalenko, I.V. \& Strong, A.W., 2000,
 Ap\&SS 272, 247 

\bibitem{berezhko} Berezhko, E.G. \& V\"olk, H.J., 2000, ApJ 540, 923

\bibitem{hill} Hill, C.T. \& Schramm, D.N., 1985,
 Phys. Rev. D31, 564

\bibitem{berezinsky} Berezinsky, V.S. \& Grigor'eva, S.T., 1988,
 Astron. Astrophys. 199, 1

\bibitem{sigl} Sigl, G., et al., 1994, Astropart. Phys. 2, 401

\bibitem{aharonian2} Aharonian, F.A., et al., 1992,
 Phys. Rev. D46, 4188

\bibitem{halzen} Halzen, F., et al., 1990, Phys. Rev. D41, 342

\bibitem{protheroe} Protheroe, R.J. \& Stanev, T., 1996,
 Phys. Rev. Lett. 77, 3708

\bibitem{sigl1} Sigl, G., et al., 1996, astro-ph/9604093 v2

\bibitem{nikolsky} Nikolsky, S.I., et al., 1987, J. Phys. G13, 883

\bibitem{suga} Suga, K., et al., 1988, ApJ 326, 1036

\bibitem{drees} Drees, M. \& Halzen, F., 1988, Phys. Rev. Lett.
 61, 275

\bibitem{kakimoto} Kakimoto, F., et al., 1991, Proc. 22nd ICRC,
 Dublin, vol. 1, 412

\bibitem{denninghoff} Denninghoff, S., PhD thesis, 2001, 
Technische Universit\"at M\"unchen

\bibitem{he} He, Y.D.\& Zhu, R.Q., 1991, Phys. Rev. D44, R2635

\bibitem{sasano} Sasano, M., et al., 1999, Proc. 26th ICRC, Salt Lake City, 
OG 2.4.14

\bibitem{karle1} Karle, A., et al., 1995, Phys. Lett. B347, 161

\bibitem{schmele} Schmele, D., PhD thesis, 1998, Universit\"at Hamburg

\bibitem{merck} Merck, M., PhD thesis, 1993,
 Ludwigs-Maximilians-Universit\"at M\"unchen

\bibitem{karle} Karle, A., et al., 1995, Astropart. Phys. 3, 321

\bibitem{moralejo} Moralejo, A., PhD thesis, 2000, Universidad
 Complutense de Madrid

\bibitem{nishimura0} Nishimura, J., 1967,
 Handbuch der Physik XLVI/2, ed. S. Fl\"ugge (Springer, Berlin) p. 1

\bibitem{capdevielle} Capdevielle, J.N., et. al., 1992, KfK Report
 4998, Institut f\"ur Kernphysik, Karlsruhe

\bibitem{martinez} Mart\'{\i}nez, S., et al., 1995, Nucl. Instr. \&
 Meth. A357, 567

\bibitem{wiebel} Wiebel-Sooth, B., et al., 1998, A\&A 330, 389

\bibitem{patterson} Patterson, J.R., \& Hillas, A.M., 1983,
 J.Phys.G 9, 1433

\bibitem{arqueros} Arqueros, F., et al., 1996,
 Astropart. Phys. 4, 309

%\bibitem{lindner} Lindner, A., 1998, Astropart. Phys. 8, 235

\bibitem{prahl} Prahl, J., PhD thesis, 1999, Universit\"at Hamburg

\bibitem{alexandreas} Alexandreas, D.E., et al., 1993, Nucl. Instr. \&
 Meth. A328, 570

\bibitem{nishimura} Nishimura, J., et al., 1980,
 Astrophys. J. 238, 394

\bibitem{musser} Musser, J.A., et al., 1997, Proc. 25th
 ICRC, Durban, 4,209

\bibitem{matthews} Matthews, J., et al., 1991, ApJ 375, 202

\bibitem{reynolds1} Reynolds, P.T., et al., 1993, ApJ 404, 206

\bibitem{aglietta1} Aglietta, M., et al., 1996,
 Astropart. Phys. 6, 71

\bibitem{chantell} Chantell, M.C., et al., 1997, Phys. Rev. Lett.
 79, 1805

\bibitem{aharonian4} Aharonian, F.A., et al., 2000, submitted
 to A\&A 

\bibitem{reynolds} Reynolds, P.T., et al., 1990, Proc. 21st ICRC,
 Adelaide, vol. 2, 383

\bibitem{aglietta} Aglietta, M., et al., 1992, ApJ 397, 148

\bibitem{amenomori} Amenomori, M., et al., 1997, Proc. 25th ICRC,
 Durban, vol. 3, 117

\bibitem{horns} Horns, D. \& Schmele, D., 1999, Proc. 26th
 ICRC, Salt Lake City, OG 3.2.24

\bibitem{borione} Borione, A., et al., 1998, ApJ 493, 175

\bibitem{bohec} LeBohec, S., et al., 2000, astro-ph/0003265


%---------------------------------------------------


%



%\bibitem{aharonian1} Aharonian, F.A., 1991, Astrophys. Space 
% Sci. 180, 305


%\bibitem{aharonian3} Aharonian, F.A. \& Atoyan, A.M., 1995,
% A\&A 294, L41




%\bibitem{almaini} Almaini, O., et al., 1999, MNRAS 305, L59



%\bibitem{barwick} Barwick, S.W., et al., 1998, ApJ 498, 779




%\bibitem{berezinsky1} Berezinsky, V.S. \& Kudryavtsev, V.A.,
% 1988, Soviet Astr. Letters 14, 370

%\bibitem{berezinsky2} Berezinsky, V.S. \& Kudryavtsev, V.A., 1990,
% ApJ 349, 620

%\bibitem{berezinsky3} Berezinsky, V.S., et al., 1993, Astropart.
% Phys. 1, 281





%\bibitem{broadbent} Broadbent, A., et al., 1989, MNRAS 237, 381



%\bibitem{chi} Chi, X., et al., 1993, AP1, 239



%\bibitem{comastri} Comastri, A., et al., 1995, A\&A 296, 1




%\bibitem{dixon} Dixon, D.D., et al., 1998, astro-ph/9803237 v2


%\bibitem{dwek} Dwek, E., et al., 1998, ApJ 508, 106

%\bibitem{fichtel} Fichtel, C., et al., 1972, ApJ 171, 31

%\bibitem{fichtel1} Fichtel, C., et al., 1975, ApJ 198, 163 

%\bibitem{gould} Gould, R.J. \& Schreder, G., 1966, Phys. Rev. Lett.
% 16, 252




%\bibitem{hasinger} Hasinger, G., 1996, A\&AS 120C, 607



%\bibitem{henry} Henry, R.C., 1991, ARA\&A 29, 89

%\bibitem{henry1} Henry, R.C. \& Murthy, J., 1994, Extragalatic 
% Background Radiation, ed. Calzetti, D., et al., Cambridge Univ. Press, 51


%\bibitem{hill1} Hill, C.T., et al., 1987, Phys. Rev. D36, 1007






%\bibitem{kamata} Kamata, K., et al., 1968, Canadian Journal of
% Phys. 46, S72





%\bibitem{kraushaar} Kraushaar, W., et al., 1972, ApJ 177, 341

%\bibitem{krawczynski} Krawczynski, H., et al., 1996, Nucl. Instr. \&
% Meth. A383, 431

%\bibitem{lawson} Lawson, K.D., et al., 1987, MNRAS 225, 307





%\bibitem{malkan} Malkan, M.A. \& Stecker, F.W., 1998, ApJ 496, 13

%\bibitem{mannheim} Mannheim,....




%\bibitem{mayer-hasselwander} Mayer-Hasselwander, H., et al., 1982,
% A\&A 105, 164

%\bibitem{cammon} McCammon, D. \& Sanders, W.T., 1990,
% ARA\&A 28, 657





%\bibitem{moskalenko} Moskalenko, I.V. \& Strong, A.W., 1998,
% ApJ 493, 694









%\bibitem{ostriker} Ostriker, J.P., et al., 1986, Phys. Lett. B180,
% 231







%\bibitem{primack} Primack, J.R., et al., 1999, ApP 11, 93


%\bibitem{protheroe1} Protheroe, R.J. \& Johnson, P.A., 1996,
% Astropart. Phys. 4, 253

%\bibitem{protheroe2} Protheroe, R.J. \& Biermann, P.L., 1996,
% ApP 6, 45

%\bibitem{protheroe3} Protheroe, R.J. \& Meyer, H., 2000, submitted to
% Phys. Lett. B, astro-ph/0005349

%\bibitem{reich} Reich, P. \& Reich, W., 1988, A\&A 196, 211




%\bibitem{sacher} Sacher, W. \& Schonfelder, V., 1983, Space Sci.
% Rev. 36, 249 
















%\bibitem{wang} Wang, Q.D. \& McCray, R., 1993, ApJ 409, L37

%\bibitem{wdowczyk} Wdowczyk, J. \& Wolfendale, A.W., 1990,
% ApJ 349, 35

%\bibitem{weekes} Weekes, T.C., 1988, Phys. Rev. 160, 1




\end{thebibliography}
\end{document}